# Scaling Up Purcell-Enhanced Self-Assembled Nanoplasmonic Perovskite Scintillators into the Bulk Regime

Michal Makowski,[1, a)] Wenzheng Ye,[2, 3] Dominik Kowal,[1] Francesco Maddalena,[2, 3] Somnath Mahato,[1] Yudhistira Tirtayasri Amrillah,[4] Weronika Zajac,[1, 5] Marcin Eugeniusz Witkowski,[6] Konrad Jacek Drozdowski,[6] Nathaniel,[4] Cuong Dang,[2, 3] Joanna Cybinska,[1, 5] Winicjusz Drozdowski,[6] Ferry Anggoro Ardy Nugroho,[4, 7] Christophe Dujardin,[8, 9] Liang Jie Wong,[2, 3, b)] and Muhammad Danang Birowosuto[1, c)]

[1)] *Lukasiewicz Research Network - PORT Polish Center for Technology Development, Wroclaw, 54-066, Poland*

[2)] *CINTRA (CNRS-International-NTU-THALES Research Alliance), IRL 3288 Research Techno Plaza, 50 Nanyang Drive, Border X Block, Level 6, Singapore 637553, Singapore*

[3)] *School of Electrical and Electronic Engineering, Nanyang Technological University, Singapore 639798, Singapore*

[4)] *Department of Physics, Faculty of Mathematics and Natural Sciences, Universitas Indonesia, Depok 16424, Indonesia*

[5)] *Faculty of Chemistry, University of Wroclaw, Wroclaw, 50-383, Poland*

[6)] *Institute of Physics, Faculty of Physics, Astronomy, and Informatics, Nicolaus Copernicus University in Torun, Torun, 87-100, Poland*

[7)] *Institute for Advanced Sustainable Materials Research and Technology (INA-SMART), Faculty of Mathematics and Natural Sciences, Universitas Indonesia, Depok 16424, Indonesia*

[8)] *Universite Claude Bernard Lyon 1, Institut Lumiere Matiere UMR 5306 CNRS, 10 rue Ada Byron, Villeurbanne, 69622, France*

[9)] *Institut Universitaire de France (IUF), 1 rue Descartes, Paris Cedex 05, 75231, France*

(Dated: 13 May 2025)

Scintillators convert high-energy radiation into detectable photons and play a crucial role in medical imaging and security applications. The enhancement of scintillator performance through nanophotonics and nanoplasmonics, specifically using the Purcell effect, has shown promise but has so far been limited to ultrathin scintillator films because of the localized nature of this effect. This study introduces a method to expand the application of nanoplasmonic scintillators to the bulk regime. By integrating 100-nm-sized plasmonic spheroid and cuboid nanoparticles with perovskite scintillator nanocrystals, we enable nanoplasmonic scintillators to function effectively within bulk-scale devices. We experimentally demonstrate power and decay rate enhancements of up to (3.20 ± 0.20) and (4.20 ± 0.31) folds for plasmonic spheroid and cuboid nanoparticles, respectively, in a 5-mm thick $CsPbBr_3$ nanocrystal-polymer scintillator at RT. Theoretical modeling also predicts similar enhancements of up to (2.26 ± 0.31) and (3.02 ± 0.69) folds for the same nanoparticle shapes and dimensions. Moreover, we demonstrate a (2.07 ± 0.39) fold increase in light yield under $^{241}$Am $\gamma$-excitation. These findings provide a viable pathway for utilizing nanoplasmonics to enhance bulk scintillator devices, advancing radiation detection technology.

[a] Electronic mail: michal.makowski@port.lukasiewicz.gov.pl  
[b] Electronic mail: liangjie.wong@ntu.edu.sg  
[c] Electronic mail: muhammad.birowosuto@port.lukasiewicz.gov.pl

# I. INTRODUCTION

The advancement of high-performance scintillating materials has remained a priority due to their applications in bioimaging[1] and various industrial technologies[2]. Following Roentgen's discovery of X-rays[3], researchers have pursued faster and brighter scintillators by investigating new materials[4] and improving crystal quality. These advances often involve modifying the properties of the emitting center using lanthanide[5] and transition metal activators[6]. More recently, nanoscale scintillators, particularly nanocrystals (NCs), have gained attention, as they offer unique pathways to improve luminescent yield and decay times at room temperature (RT) compared to their bulk associates[7,8]. However, these intrinsic material properties alone are not the sole determinants of scintillator performance. Increasingly, nanophotonic and nanoplasmonic techniques[9–11], such as the Purcell effect, are being explored to further enhance emission characteristics by modifying the local density of optical states (LDOS)[11–16]. By optimizing the LDOS, the Purcell effect can significantly amplify radiative decay rates, potentially increasing the scintillation efficiency.

Whereas single-crystal scintillators such as Cerium-doped Lutetium Oxide and Lanthanum Bromide, Thallium-doped Sodium and Cesium Iodides, and Bismuth Germanate are commonly used in medical imaging applications operating in a counting regime, including positron emission tomography (PET)[17], the class of perovskite scintillators has emerged as promising materials in their own right[18–20]. Perovskites offer several advantages that make them attractive alternatives to traditional single-crystal scintillators. Notably, perovskite light yields are within tens of photons per keV (ph/keV), comparable to or surpassing those of established materials. In addition, perovskites exhibit fast decay times of less than 10 nanoseconds, enhancing the timing resolution of the imaging system. The fabrication processes for perovskite materials are often more straightforward and, thus, more cost-effective than those for traditional single-crystal scintillators. Finally, the tunability of perovskite materials enables the optimization of their scintillation properties through compositional adjustments, potentially beckoning the emergence of novel scintillator designs tailored for specific imaging requirements[11]. However, like other nanophotonic methods used to enhance the properties of the scintillator, the Purcell enhancement of plasmonic thin films yield is limited to a certain thickness of the scintillator material. This phenomenon occurs due to the typically localized nature of the Purcell effect. The challenge lies in determining how to scale up these enhancements for bulk scintillators, whose thicknesses can be as large as millimeters in scale[11,13–16].

This work overcomes this challenge by demonstrating the potential for nanoplasmonic Purcell enhancement in bulk scintillator materials, achieved by embedding $CsPbBr_3$[21,22] perovskite NCs with silver (Ag) nanoparticles (NPs) in a polydimethylsiloxane (PDMS) matrix, creating stable, bright, and scalable scintillating materials suitable for medical and security imaging applications. Although our approach employs silver, which belongs to precious metals, it remains significantly cheaper than commonly used scintillators, with $LaBr_3$:Ce being a prime example. Production costs of 1 cm$^3$ of $LaBr_3$:Ce can reach 750 euros, while our approach reduces this price to 90 euros/cm$^3$. Moreover, Ag is selected because it can provide the highest enhancements and the lowest losses for the green emission of our NCs[11]. This hybrid system leverages the Purcell effect, in which Ag NPs modify the LDOS, amplifying the light yields and decay rates of $CsPbBr_3$ NCs without thickness limitations. The concept from nano to bulk was also demonstrated by Förster-resonance-energy-transfer $CsPbBr_3$ scintillators[23], but our concept is more robust for large areas while providing extensive enhancements. By tuning the interaction between Ag spheroid NPs (SNPs) and cuboid NPs (CNPs) and $CsPbBr_3$ NCs, this work proposes an original route based on two configurations to improve scintillation in large-scale systems, achieving intensified luminescence and faster decay dynamics than previously reported thin-film configurations[11,13–16].

The Purcell effect in our experiment is determined by comparing the luminescence intensity (emitted power, $P$) and decay time (inverse decay rate, $\Gamma$) ratios between $CsPbBr_3$ NCs dispersed within Ag NP-PDMS composite systems and those in pure PDMS matrices, denoted by a superscript $^o$. The Purcell factor $F_p$, describing this enhancement, is defined as[24]:

$$F_p \sim \frac{P}{P^o} \sim \frac{\Gamma}{\Gamma^o} \quad (1)$$

where the ratios $\Gamma/\Gamma^o$ and $P/P^o$ both express $F_p$, depending on the quantum efficiency ($QE$) of the $CsPbBr_3$ NCs[25,26]. To relate $F_p$ to the LDOS, one can further express the $\Gamma/\Gamma^o$ ratio using a classical dipole model of the electric Green dyadic $\mathbb{G}^{\mathbb{E}}(\mathbf{r_s}, \mathbf{r'_s}, \omega_{eg})$:

$$\frac{\Gamma}{\Gamma^o} = \frac{6\pi c}{\omega_{eg}} \text{Im}\left[\mathbf{u} \cdot \mathbb{G}^{\mathbb{E}}(\mathbf{r_s}, \mathbf{r_s}, \omega_{eg})\right] \quad (2)$$

where $\omega_{eg}$ is the Bohr frequency, $\mathbf{u}$ is the unit vector along the direction of the dipole, $\mathbf{r_s}$ is the emitter position, and $c$ is the speed of light. To compute $P/P^o$, a substitution of $\omega_{eg}$ with the frequency of one of the eigenmodes $\omega$ is employed[24].

Although the Ag NP systems used here are randomly distributed, their interactions can be approximated by modeling a single NP if all NCs and Ag NPs are 100% coupled together or the coupling efficiency ($C_{coup}$) is 100%. Multiple scattering also minimally influences average decay rate statistics in these disordered systems if the scattering strength is low enough due to the lower index contrast between medium (PDMS) and scatterers (Ag NPs) than between air and Ag NPs[27,28]. The multiple scattering approximation affects only the distribution tail[29], which is not present in our experiments or the previous model of ensemble Förster-resonance-energy-transfer scintillator systems[23]. Thus, the Purcell factor of our scintillators can be derived using previous studies on superemitters, hybrid emitter systems, and plasmonic NPs[30,31]. For a single CsPbBr$_3$ NC coupled to a single Ag NP, $F_p$ is given by[32]:

$$F_p = \frac{3}{4\pi^2}\lambda_{eg}^3 Q\left(\frac{1}{V}\right) = \frac{6\pi c^3}{\omega_{eg}^3}Q\left(\frac{\max(\epsilon|E|^2)}{\int \epsilon|E|^2 d^3\mathbf{r}}\right) \qquad (3)$$

where $\lambda_{eg} = 2\pi c/\omega_{eg}$ and $Q$ is the quality factor of the Ag NP. The parameters $Q$ and the volume of the mode $V$ depend on the details of the NPs, such as SNPs and CNPs, as solved analytically in the Supporting Information and Figure S1 (Supporting Information). For plasmonic $F_p$ calculations, $V$ has a greater impact than $Q$ due to the typically low $Q$ values (under ten)[32]. To determine $V$, the integral over the energy density is calculated, $(\text{Re}\{\epsilon\} + 2\omega\text{Im}\{\epsilon\}\gamma)|E|^2$ [33]. This approach fits well with the Drude model $\epsilon = 1 - \omega_p^2/\omega(\omega + i\gamma)$, where $\omega_p$ and $\gamma$ denote the frequency of the plasmonic mode and the damping rate for Ag, respectively[32]. Because FDTD simulations directly compute the ratio of radiated power in the presence and absence of nanostructures, the resulting Purcell factor aligns with Eq. 3, which is derived from Eq. 1 through an integral on the energy density. According to Eq. 3, $F_p$ is strongly influenced by the maximum energy density in the integral, making it responsive to sharp NP geometries[33]. Structures like CNPs, which feature sharp edges, achieve significantly higher $F_p$ values than SNPs due to these geometric effects[34–36].

This study examines the intricate interactions between CsPbBr$_3$ NCs and Ag NPs in a PDMS matrix, focusing on the impact of Ag NP geometry on Purcell enhancements. The analysis integrates experimental findings with theoretical insights and FDTD simulations, incorporates corrections for $QE$ and $C_{coup}$ based on emitter and microscopy characterizations, and compares theoretical predictions with experimental results. Photoluminescence (PL) studies reveal an integrated intensity enhancement of $P_{PL}/P_{PL}^o$ at (4.10 ± 0.20) folds at RT. Furthermore, time-resolved PL (TRPL) measurements show a significant reduction in the fastest decay time constant, dropping from (0.99 ± 0.03) ns to (0.61 ± 0.04) ns, producing a maximum overall decay rate ratio $\Gamma_{PL}/\Gamma_{PL}^o$ of (4.20 ± 0.31) folds. For X-ray luminescence (XL), the maximum enhancements reach (1.92 ± 0.13) folds in $P_{XL}/P_{XL}^o$ and (2.08 ± 0.06) folds in $\Gamma_{XL}/\Gamma_{XL}^o$. Under 59.5 keV $\gamma$ excitation from the $^{241}$Am source a (2.07 ± 0.39) enhancement of $P_\gamma/P_\gamma^o$ was measured. These values align closely with the theoretical enhancements, and CNP-doped samples achieve the highest values. This confirms the origin of the Purcell effect while highlighting the limitations of observational methods[11]. Crucially, our Purcell-enhanced scintillator system is designed with a millimeter-scale thickness, addressing the constraints of thin-layer designs. This advancement paves the way for its application in high-energy radiation fields, such as PET and photon-counting computed tomography (PCCT)[37].

## II. RESULTS

The successful demonstration of the Purcell effect with a single plasmonic nanoparticle critically depends on the precise positioning of the emitter and maintaining proximity to the nanoparticle, without direct contact with its metal surface. High-quality plasmonic NPs are commonly coated with a 5 nm thick poly(vinylpyrrolidone) (PVP) layer acting as a surfactant, facilitating various light-matter interactions, from weak to strong coupling[38,39]. Ag NPs were selected over gold (Au) NPs due to their substantially lower optical losses, which prior studies have shown to produce superior Purcell enhancements in Ag-based nanoplasmonic systems, compared to Au-based systems[11]. The schematic diagrams of the NC and NC-NP scintillators are presented in Figs. 1a, b and c, respectively. Transmission Electron Microscopy (TEM) was used to assess the size of Ag NPs, revealing diameters of (100 ± 8) nm for SNPs and side lengths of (100 ± 10) nm for CNPs (Figs. 1d and e, respectively). The white dashed lines represent the 5 nm thick PVP shielding, which, due to size limitations, are hardly visible under TEM. Thicker silica coatings are well visible with TEM imaging as presented by the manufacturer[40]. Figure 1f shows an image of the synthesized samples after PDMS fabrication under ambient conditions, while Figure 1g depicts the samples under UV lamp excitation. Although the side surfaces of the fabricated cylindrical samples appear slightly rough to the eye, they are not critical in the frame of the scintillation purpose. At first sight, the sample with CNPs is the brightest. For their dimension, the samples were exhibited with a diameter $\phi$ of 10 mm and a thickness of 5 mm. Theoretically, such thickness is adequate for all samples to fully absorb high-energy radiation within the range of 10 to 50 keV; see Figure S2 (Supporting Information).

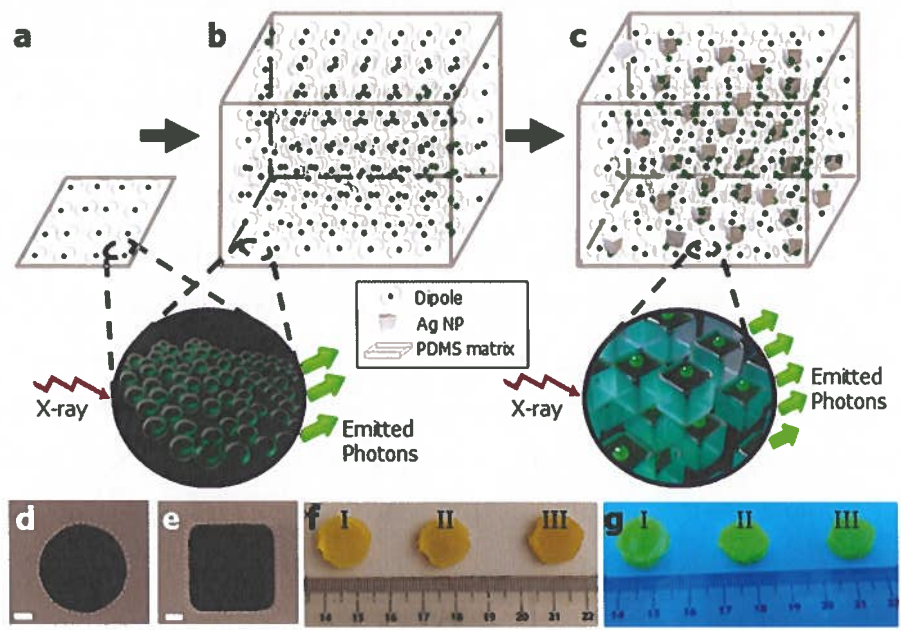

FIG. 1. **Nanoplasmonic scintillators with metal nanoparticles. a,b,c,** Schematic descriptions of available $CsPbBr_3$ nanocrystals (NCs) thin films (**a**), our concept of $CsPbBr_3$ NCs (**b**) and $CsPbBr_3$ NCs sample co-doped with silver nanoparticles (Ag NPs) (**c**) embedded into polydimethylsiloxane (PDMS) matrix. The replication and the shortening of the light green arrows indicate a scintillation light yield increase and a decay time shortening, respectively. The sample with NPs shows brighter NCs inside the matrix. **d,e,** High-resolution transmission electron microscopy (TEM) images of individual spheroid (SNPs) (**d**) and cuboid nanoparticles (CNPs) (**e**) with 5 nm thick PVP coating layer outlined with white dashed lines (white scale bar: 20 nm). **f,g,** Images of the synthesized samples captured under ambient conditions (**f**) and UV lamp excitation. (**g**), highlighting the distinct brightnesses of $CsPbBr_3$ + SNPs (**I**), and $CsPbBr_3$ + CNPs (**II**), and pure $CsPbBr_3$ (**III**).

To visualize the attachment of CsPbBr$_3$ NCs with Ag NPs and to investigate the shape and size of individual NCs, atomically resolved TEM measurements were performed (Figure 2). Considering the highly beam-sensitive nature of perovskites, the CsPbBr$_3$ NCs are defined by low-dose aberration-corrected scanning TEM high-angle-annular dark-field (STEM-HAADF) imaging operated at 300 kV with a screen current of ∼20 pA[41]. As can be seen in the low magnification TEM images, the CsPbBr$_3$ NCs are well attached to the Ag NPs and uniformly distributed on the Ag NP surface (Figure 2a). The average side lengths of the CsPbBr$_3$ NCs are assessed to be (7.30 ± 0.20) nm and (6.76 ± 0.20) nm while the atomic distance of Pb-Pb atoms is 0.59 nm (Figure 2b and Figure S3, Supporting Information). To verify the crystallinity and growth of NCs, the Fast Fourier Transform patterns derived from Figure 2b are shown in Figure 2c. The well-defined diffraction spots indicate the high crystallinity of CsPbBr$_3$ NCs. The indexed diffraction spots of the crystal planes (100) and (110) along the [001] zone axis also indicate the cubic phase in CsPbBr$_3$[42]. Subsequently, to visualize Ag NPs, the well-resolved lattice fringes (Figure 2d) with an interplanar spacing of 0.40 nm correspond to the (100) crystal faces of Ag-Ag atoms (Figure S3, Supporting Information)[43]. Verifying the chemical composition at the atomic scale can provide direct evidence for determining the materials' structures and chemical information, and this has been demonstrated with energy-dispersive X-ray spectra (EDS). The region for EDS spectral imaging is shown in Figure 2e with the STEM-HAADF image. Elemental maps shown in Figures 2f-i were extracted from EDS spectral images with selected energy windows for Br, Cs, Pb, and Ag, respectively. Consequently, calculations were performed to estimate the amount of CsPbBr$_3$ NCs coupled to the surfaces of Ag NPs. EDS images showing the overlap between Ag and Pb or Br elements were analyzed using Sørensen-Dice similarity method[44], and the coupling percentage for both Ag NP-doped samples is estimated to be $C_{coup}$ = (70 ± 8) % (Figure S3, Supporting Information). In Figure S3 (Supporting Information), the analysis of the shortest distance between NC atoms and Ag NP atoms in the EDS images yields values comparable to the 5 nm thickness of the PVP layer, accounting for potential polymer compression.

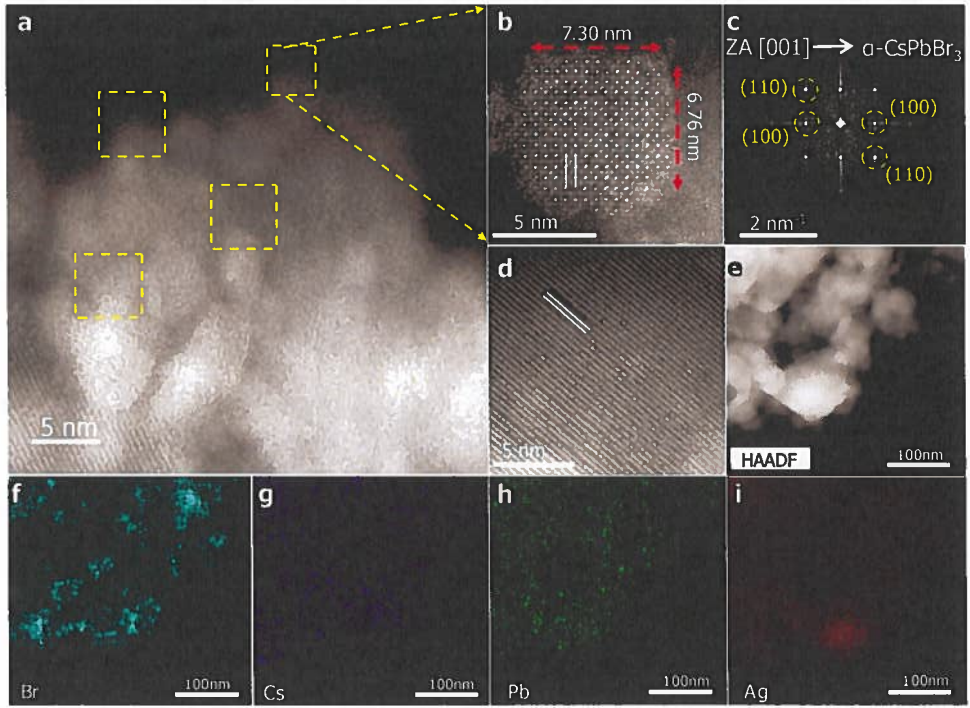

FIG. 2. **Morphology of Ag CNPs surrounded by myriad CsPbBr$_3$ NCs. a**, Scanning TEM high-angle annular dark-field (STEM-HAADF) image of the Ag CNP embedded within the CsPbBr$_3$ NCs. Areas marked with yellow squares indicate regions selected for energy-dispersive X-ray analysis. **b,c**, Atomic resolution STEM-HAADF image with the lattice spacing of 0.59 nm shown as white parallel lines (**b**) of the CsPbBr$_3$ NCs, with corresponding Fast Fourier Transform image (**c**) revealing diffraction spots indexed with (100) and (110) planes, confirming the cubic phase of CsPbBr$_3$. **d**, High-resolution STEM-HAADF image with the lattice spacing of 0.40 nm from the corresponding Ag atoms shown as white parallel lines. **e**, A low magnification STEM-HAADF image of Ag embedded with CsPbBr$_3$ NCs. **f-i**, X-ray elemental maps for Br (**f**), Cs (**g**), Pb (**h**), and Ag (**i**), showcasing the compositional distribution in this hybrid system.

To clarify the role of Ag NPs in enhancing emitter characteristics and to validate our Purcell enhancement strategy, the FDTD simulations for CsPbBr$_3$ NCs attached to a single Ag NP were performed. These simulations were arranged with a 9 nm emitter-NP separation (corresponding to the thickness of the PVP layer of 5 nm and the radius of CsPbBr$_3$ NCs), an isotropic dipole orientation[45], an SNP diameter of 100 nm, a CNP side length of 100 nm and PDMS medium. For CNP, the edges and corners of silver CNP were rounded to 3 nm and 5 nm, respectively, according to previous microscope image studies[38]. Changing the shape of Ag NPs, an effect on emitter decay rates was analyzed (Figs. 3a and b). Our simulations further suggest that the embedding of single CNPs with single CsPbBr$_3$ NCs in a PDMS matrix can enhance LDOS by up to **45** times, particularly near the edges or corners of the CNP compared to SNP. The complete set of equations for this analysis is provided in the Supporting Information.

Following an initial assessment of single-emitter interactions, the analysis was extended to systems with multiple emitters in single Ag NPs. Using Monte Carlo simulations, 100,000 configurations with random emitter distributions based on the estimated surface coverage in the samples were generated: 68 and 126 emitters for SNPs and CNPs, respectively (see Methods). Due to the substantial number of configurations, nonlocal optical responses[46] were not included in our calculations, as these effects are averaged statistically when using the ratio definitions in Eq. 1. A histogram analysis of the enhancement of LDOS in all configurations indicates that CNPs produce a more significant enhancement than SNPs (Figs. 3c and d). Specifically, SNPs achieve an average enhancement factor of (5.87 ± 0.01), while CNPs reach (7.85 ± 1.80). The SNPs were expected to show no variations as a result of their spheroid shape and the absence of sharp edges. The minor variation in the histogram for the SNPs is due to the limitation of the mesh to form a round shape for the calculation, although the mesh size is already 0.5 nm. Finally, these values will be corrected for the $QE$ of CsPbBr$_3$ NC and $C_{coup}$ of our multiple NC-NP systems. Furthermore, the observed enhancement is likely influenced by surface effects of both CsPbBr$_3$ NCs and Ag NPs. Surface defects in NCs can contribute to nonradiative recombination, but this is accounted for by using the measured $QE$ of 55% in our calculations. Additionally, the PVP buffer layer on the Ag NP surface maintains a 5 nm separation from the NCs, mitigating direct quenching while still allowing significant plasmonic enhancement.

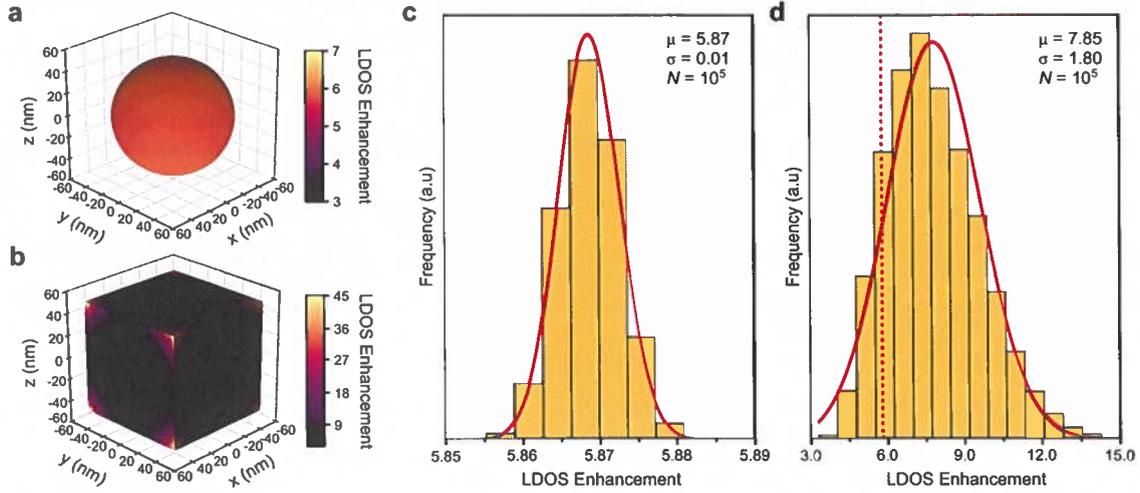

FIG. 3. **Theoretical calculations.** **a**, **b**, Three-dimensional maps of local density of states (LDOS) enhancements for CsPbBr$_3$ NCs doped with SNPs (**a**) and CNPs (**b**), respectively. **c**, **d**, The histograms of LDOS enhancements for SNP- (**c**) and CNP-doped (**d**) samples. Red dashed line in **d** represents the normal distribution from **c**, to show the overlap with the distribution tail. The histograms represent 10,000 random configurations of 68 and 126 emitters for SNPs and CNPs, respectively.

Experiments to probe the Purcell enhancements were then performed and analyzed for two sample points each. Optical characterizations were initially performed through PL and TRPL measurements. Before determining the Purcell enhancements from both measurements, measuring the CsPbBr$_3$ NC $QE$ using an integrating sphere was necessary. From the series of measurements, the average value $QE$ was $(55 \pm 15)\%$ (Figure S4, Supporting Information). With this $QE$, it is expected to see moderate enhancements in integrated luminescence intensities and decay rates[11,25]. Although there are reports on CsPbBr$_3$ showing values of $QE$ close to unity, the Purcell effect would still increase $\Gamma_{rad}$, reducing overall decay while maintaining $P_{rad}$. This is crucial for Time-of-Flight (TOF) applications such as those for PET. Figure 4a shows the RT PL spectra, where samples doped with Ag NPs exhibit significantly higher intensities compared to pure CsPbBr$_3$ samples. Specifically, the integrated PL intensity enhancements for SNPs ($P_{PL}^{SNP}/P_{PL}^o$) and CNPs ($P_{PL}^{CNP}/P_{PL}^o$) are $(3.20 \pm 0.20)$ and $(4.10 \pm 0.20)$ folds, respectively. Figure 4b presents the RT TRPL decay curves, with both Ag-doped samples showing faster PL decays. At first sight, two decay components are related to multiexciton generations in CsPbBr$_3$ NCs[8,47]. Using biexponential decay fitting (Supporting Table S1), the shortest decay time constants decrease from $(0.99 \pm 0.03)$ ns in pure CsPbBr$_3$ to $(0.71 \pm 0.05)$ ns and $(0.61 \pm 0.04)$ ns in SNP- and CNP-doped samples, respectively. Both values are slower than 0.3 ns of the instrument response function (IRF), see Methods. The average PL decay times ($\bar{\tau}_{PL}$) at RT are $(4.02 \pm 0.13)$ ns, $(1.61 \pm 0.08)$ ns, and $(0.94 \pm 0.06)$ ns for pure, SNP-, and CNP-doped samples, respectively, resulting in PL decay rate enhancements of $(2.48 \pm 0.17)$ and $(4.20 \pm 0.31)$ fold for SNP- ($\Gamma_{PL}^{SNP}/P_{PL}^{o}$) and CNP-doped samples ($\Gamma_{PL}^{CNP}/P_{PL}^{o}$), respectively.

Subsequent to the PL and TRPL measurements, XL and TRXL characterizations were performed, anticipating similar enhancements in integrated luminescence and decay rate[11]. As shown in Figure 4c, XL spectra reveal enhancements of $(1.71 \pm 0.09)$ and $(1.92 \pm 0.13)$ folds for SNP- ($P_{XL}^{SNP}/P_{XL}^{o}$) and CNP- ($P_{XL}^{CNP}/P_{XL}^{o}$) doped samples, respectively. RT TRXL decay curves (Figure 4d) reveal a triexponential decay with a notable long component (> 10 ns), detailed in Table S1 (Supporting Information). The additional component appears in TRXL decay curves due to the higher energy excitation of TRXL compared to that of TRPL, generating multiple intermediate states, including trap defect states, and involves multiexcitonic processes[8,47]. The fastest decay time is still that of CNPs with $(0.67 \pm 0.04)$ ns, which is still slower than IRF, see Methods. The average decay times $\bar{\tau}_{XL}$ decrease from $(7.55 \pm 0.12)$ ns in the pure sample to $(4.18 \pm 0.13)$ ns and $(3.70 \pm 0.10)$ ns in the respective SNP- and CNP-doped samples, resulting in enhancements in the XL decay rates, $\Gamma_{XL}^{SNP}/P_{XL}^{o}$ and $\Gamma_{XL}^{CNP}/P_{XL}^{o}$, of $(1.84 \pm 0.07)$ and $(2.08 \pm 0.06)$ folds, respectively. These enhancements correlate with surface plasmon resonance effects, which increase the LDOS near the NPs, favoring higher radiative recombination probabilities. CNPs demonstrated the strongest enhancements presumably due to their intense and narrow scattering spectra, optimizing overlap with the CsPbBr$_3$ NC emission spectrum (Figure S2, Supporting Information). Given the enhancements in both the light yield and decay rates, future investigations will focus on coincidence measurements, as CNP doping could reduce the overall coincidence time resolution (CTR $\propto \sqrt{\tau/\text{LY}}$) by 50% compared to pure CsPbBr$_3$, which is crucial for PET applications[47].

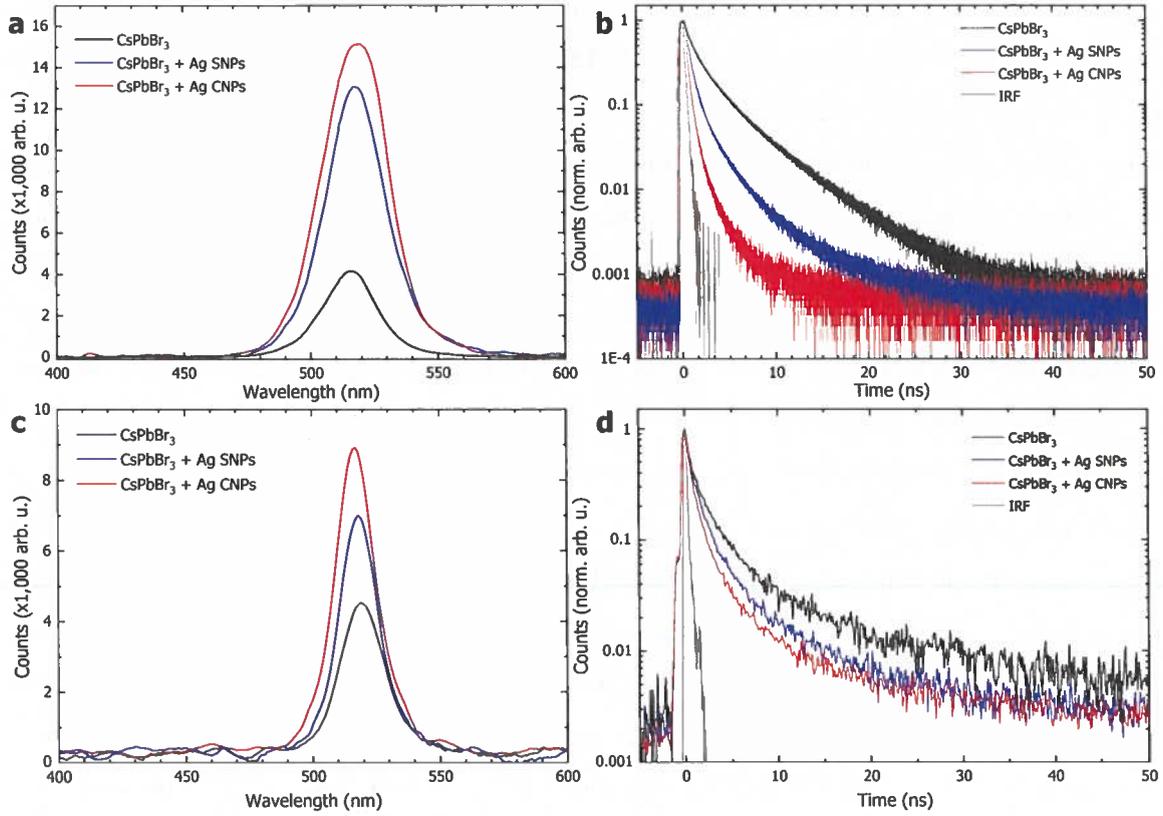

FIG. 4. **Optical and X-ray excitation characterizations. a, b** The photoluminescence (PL) spectra (**a**) and time-resolved PL decay curves (**b**). **c, d** The X-ray luminescence (XL) spectra (**c**) and time-resolved XL decay curves (**d**). All measurements were performed at room temperature (RT). Notably, the PL spectrum and the decay curve for $CsPbBr_3$ + CNPs exhibit the most significant plasmonic enhancement, reaching about 4-fold, while other enhancements demonstrate approximately between 2- and 3-fold enhancements.

To evaluate potential enhancements in $QE$ at low temperature (LT), we conducted temperature-dependent PL and XL measurements (Figure S5, Supporting Information). For the pure sample, there is a very small redshift in the XL spectra at 10 K of about 3 nm due to exciton-phonon interaction in $CsPbBr_3$ NCs[47]. Remarkably, for CNP-doped samples, a much larger 10-nm redshift in emission peak wavelength is observed at LT, with the XL peak shifting from 518 nm at RT to 528 nm at 10 K, bringing it closer to the 541-nm peak of the CNP scattering spectrum (Figure S2, Supporting Information). The significant change in CNPs strongly suggests an enhanced light-matter coupling between $CsPbBr_3$ NC Ag CNP emissions[38,39]. Furthermore, PL and XL intensities increase in LT following the same quenching behavior (see Tables S2 and S3 for parameters, Supporting Information) and indicate a reduced contribution from nonradiative recombination processes. This directly implies a boost $QE$ and, consequently, elevated Purcell enhancements; see Supporting Information[26]. For example, at 80 K, the integrated PL intensity ratios $P_{PL}^{SNP}/P_{PL}^o$ and $P_{PL}^{CNP}/P_{PL}^o$ increase to $(5.74 \pm 0.34)$ and $(6.52 \pm 0.41)$, respectively. This increase in $QE$ in LT is also supported by the absence of traps in thermoluminescence (TL) measurements (Figure S6, Supporting Information), which is advantageous for scintillator applications that are fast and accurate in detecting ionizing radiation[1]. Furthermore, the absence of afterglow at 10 K explains why the enhancements under X-ray excitation are clearly observed.

Building on the observed improvements, X-ray imaging (Figs. 5a-c) reveals that Ag NP-doped samples demonstrate enhanced brightness, with maximum intensity measurements reaching 1,000 counts for pure CsPbBr$_3$, 1,950 counts for SNP-doped, and 2,250 counts for CNP-doped materials. This results in notably more precise imaging than previously reported in plasmonic scintillator[11]. Detailed cross-sectional images and resolution card imaging are presented in Figure S7 and Table S4 (Supporting Information). The best spatial resolution is 5 line pairs per millimeter at the 0.2 modulation transfer function, similar to that observed by Maddalena et al.[47]. However, this study focuses on Purcell enhancements in X-ray imaging intensities, where $P_{Img}^{SNP}/P_{Img}^o$ and $P_{Img}^{CNP}/P_{Img}^o$ reach enhancements of $(1.95 \pm 0.10)$ and $(2.25 \pm 0.10)$ folds, respectively. Additionally, such an NC-NP configuration may facilitate directed emission in CNP-doped materials, which can be aligned during fabrication[48].

To evaluate the enhancement under $\gamma$-irradiation and further prove the bulk nature of the observed Purcell enhancement, pulse height spectrum measurements were performed using 59.5 keV $\gamma$-rays from the $^{241}$Am source. However, in most of the trials, the photopeaks remained inconspicuous. The separation of a photopeak strongly depends on the scattering of random NCs and NC loading[49]. The appearance of a photopeak is often blurred and hidden in the measurement background. However, the extended tail observed in the pulse height spectrum of CNP-doped materials may suggest enhancement (Figure S8a and S8b, Supporting Information). With better aligned and lower loaded NCs, we obtained a partially resolved photopeak for pure CsPbBr$_3$ and for Ag CNP-doped samples. The results are presented in Figure S8c (Supporting Information). Pure CsPbBr$_3$ is characterized by a light yield of 4.1 ph/keV, while 8.5 ph/keV was measured for the Ag CNP-doped sample.

The value of our CsPbBr$_3$ nanocomposite is similar to 4,8 ph/keV reported by Erroi et al.[50]. As can be seen, a (2.07 ± 0.39) fold enhancement has been observed with co-doping with Ag CNPs. We note that the light yield values are lower than previously reported 21 ph/keV[47], but were obtained for a different type of medium. As was shown before, the choice of polymer is crucial for the evaluation of the light yield and can strongly reduce its value[51]. Unfortunately, the matrix employed in composites characterized by high light yield is subject to the company's nondisclosure policy. Despite the quenched light yield, our results are comparable to those previously reported by Zaffalon et al., where a value of up to 10 ph/keV (at 80K) was achieved[52]. Notably, their evaluation was based on integrated intensities from XL spectra. Our results were obtained under γ-excitation along with the spectral matching corrections for an absolute light yield value estimation. By applying the same procedure of XL-based light yield estimation, we obtained (7.5 ± 1.0) ph/keV for Ag CNP-doped sample; see Figure S8d (Supporting Information). All measurements consistently show the highest enhancements in samples doped with Ag CNPs (Figure 5d).

The values of FDTD simulations with a single Ag NP approach were corrected with $QE$ and $C_{coup}$ resulting in (2.26 ± 0.31) and (3.02 ± 0.69) folds for CsPbBr$_3$ NCs inside ensembles of Ag SNPs and CNPs, respectively. Interestingly, an analytical approach employs Equations S1-S11 (Supporting Information) results with perfect match with enhancement values of (2.43 ± 0.64) and (3.03 ± 0.23) for Ag SNP and CNP-doped samples, respectively. The more substantial enhancement observed for Ag CNPs in comparison to SNPs is attributed to their sharper geometric features, which generate intense plasmonic hotspots, further amplifying the LDOS. This leads to a more efficient modification of the LDOS, improving the spontaneous emission rate of CsPbBr$_3$ NCs and allowing significant enhancements of scintillation in the bulk material. Thus, we found that experimental enhancements from PL and TRPL become similar to the theoretical values, within the error margin (Figure 5d and Supporting Information). Lower absolute enhancements observed under X-ray and γ-ray excitation, compared to those observed under optical excitation, can be attributed to several factors, especially the nonproportionality between optical excitation and high energy excitation[47,53,54]. In CsPbBr$_3$ NCs[47,53], this nonproportionality with different low-energy excitations is more significant compared to two-dimensional perovskite materials[11,55]. Moreover, in PL, the yield is directly influenced by quenching and radiative probability, making the impact of Purcell enhancement straightforward. In contrast, scintillation involves multi-ple intermediate stages before reaching the final phase of light emission. Consequently, the yield is governed by various quenching phenomena throughout these stages. The Purcell effect acts solely on the final stage, reducing an overall enhancement of the entire scintillation process.

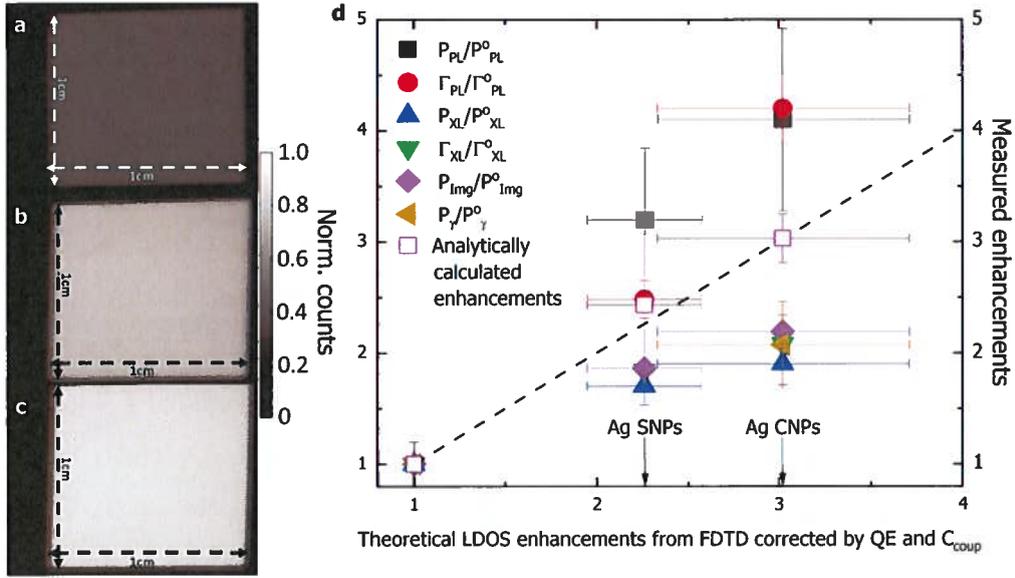

FIG. 5. **X-ray images and Purcell enhancements.** a, b, c, X-ray images of a card featuring a square hole with a side length of 1 cm for pure $CsPbBr_3$ (a), $CsPbBr_3$ + SNPs (b), and $CsPbBr_3$ + CNPs (c). d, The measured (at RT) vs theoretical LDOS enhancements, derived from FDTD calculations with correction factors: assumptions of $(55 \pm 15)$ % quantum ($QE$) and $(70 \pm 8)$ % NC-NP coupling ($C_{coup}$) efficiencies. The measured enhancements for integrated PL intensities, PL decay rates, integrated XL intensities, XL decay rates, X-ray image intensities, and $\gamma$-light yield are represented as $P_{PL}/P^o_{PL}$, $\Gamma_{PL}/\Gamma^o_{PL}$, $P_{XL}/P^o_{XL}$, $\Gamma_{XL}/\Gamma^o_{XL}$, $P_{Img}/P^o_{Img}$, and $P_\gamma/P^o_\gamma$ respectively. The empty magenta square indicates analytically calculated enhancements. The black dashed line shows a one-to-one match.

## III. DISCUSSION

This study demonstrates the first practical application of the Purcell effect to enhance X-ray and $\gamma$-ray scintillation in $CsPbBr_3$ NCs embedded in a millimeter-thick PDMS matrix. By introducing Ag SNPs and Ag CNPs, significant improvements in PL, XL, TRPL, and TRXL were observed, with the most pronounced enhancements achieved in samples doped with Ag CNPs. In RT, these samples showed enhancements up to $(3.20 \pm 0.20)$ and $(4.20 \pm 0.31)$ folds in $CsPbBr_3$ NCs with Ag SNPs and Ag CNPs, respectively. Our pure $CsPbBr_3$ sample has a light yield of 4.1 ph/keV. The observed reduction is due to the polymer coating[50–52]. Even with reduced light yield, we enhanced it by $(2.07 \pm 0.39)$ with Ag CNPs doping, reaching 8.5 ph/keV. Given that the light yields of $CsPbBr_3$ NCs in other reports are around 21 ph/keV[8,47,56], these Purcell-enhanced scintillators (after medium optimization) could exhibit significantly higher light yields (surpassing 40 ph/keV), ultrafast decay times (ranging from 0.61 to 0.67 ns), and zero afterglow, making them ideal for advanced medical imaging (e.g. PET and PCCT) and security applications[37]. To clarify the origin of observed enhancements, we note that they arise from the Purcell effect rather than the increased density of Ag NPs-doped samples. As shown in Figure S1a (Supporting Information), all X-ray and $\gamma$-ray radiation is absorbed in our 5 mm thick samples. Although the presence of Ag NPs enhances the stopping power, this effect does not influence the emission process, as Ag NPs are coated with an insulating PVP layer that prevents any charge transfer from excited NPs to $CsPbBr_3$ NCs. This conclusion is further supported by time-resolved measurements showing comparable enhancements in TRPL and TRXL. If charge transfer contributed, we would expect a prolongation of the decay time rather than the observed shortening[57]. Our work shows an alternative strategy to overcome thin film limitations[11–15] and achieves stable enhancements that are two to four times larger than those reported for nanoparticle-doped liquid scintillators[58]. Furthermore, based on a scalable self-assembly process, our millimeter-scale approach supports the emerging need for large high-performance nanophotonic scintillators[10], bypassing typical thickness limitations. Future research will prioritize optimizing nanoemitters for improved $QE$, utilizing core-shell structures to minimize nonradiative losses, exploring nanoplatelet configurations for optimal dipole alignment, and increasing Stokes shifts to reduce self-absorption[56]. Our work presents a novel and generalizable method to increase the thickness and volume of Purcell-enhanced scintillators, using CsPbBr$_3$ NCs embedded in a PDMS matrix as a representative system. Crucially, this method is not emitter-specific; it remains effective provided the scattering spectrum of the chosen metallic NP is well-matched to the emitter luminescence. Such scalability is essential for real-world applications, which often require scintillators of a certain thickness[59,60].

## IV. EXPERIMENTAL SECTION

**FDTD calculations**. Simulations were conducted using Lumerical FDTD Solutions. A simulation region span of 400 nm was used across all axes with Perfectly Matched Layer (PML) boundary conditions, a mesh size of 0.5 nm and at 300 K. The Ag NPs used the permittivity from[61], with the respective 3- and 5-nm rounded edges and corners for the CNPs. To represent the emitters in the calculations, a single broad-band dipole source was placed at the emitter position[62,63] (9 nm above the NP surface), oriented parallel or perpendicular to the NP surface. Simulations were run for these single dipoles at different positions along the Ag nanoparticle surface. The Purcell factor in each emitter position was calculated as the ratio of emitted power with and without Ag nanoparticles and was averaged from the parallel and perpendicular orientations, assuming isotropic emission[45]. The influence of Ag NP size on LDOS enhancements was calculated and is presented in Figures S9 and S10 (Supporting Information).

**CsPbBr$_3$ NCs synthesis**. NCs were prepared by a hot injection method as described in detail by Ghorai et al.[64]. The only difference was the choice of toluene as the protective medium instead of hexane for the NC solution. The side lengths of the TEM of the single CsPbBr$_3$ NC are (7.30 ± 0.20) nm and (6.76 ± 0.20) nm, see Figure 2.

**Sample preparation**. Three solutions were prepared, each containing a 20 weight percent concentration of CsPbBr$_3$ NCs. One solution was left unchanged, while the other two were supplemented with Nanocompsix Ag SNPs and Ag CNPs, respectively, at a 1:1 weight ratio of Ag to NCs. These solutions were carefully processed to promote the attachment of NC to NP[38,39]. A polymer matrix was prepared by mixing 2.7 grams of Sylgard 184 silicone elastomer (PDMS) with 0.3 grams of curing agent. Then each NC solution was added to the polymer and the mixture was agitated on a vortex mixer for 5 minutes to ensure an even distribution of CsPbBr$_3$ NCs and Ag NPs. The mixture was divided equally into three vials and cured in a vacuum oven at 120°C for 24 hours. After curing, the samples were removed from the mold, resulting in smooth surfaces with slightly rough edges, as illustrated in Figure 1a. Each sample measured 1.5 cm in diameter and 0.5 cm in thickness. Transmission spectra were measured to evaluate the transparency of prepared samples; see Supporting Figure S11 (Supporting Information). We show that even for a 12 mm thick, the integrated transmission over luminescence of the sample exceeded 50%. Experimentally, the density of perovskite NCs in the matrix is estimated to be around $1.4 \cdot 10^{13}$. The coverage of NCs on single SNP and CNP is estimated at around 68 and 126, respectively. The density of SNPs and CNPs is calculated to be around $2 \cdot 10^{11}$ and $10^{11}$, respectively.

**PL and TRPL measurements.** The initial phase of our experimental procedures to assess PL involved employing a picosecond laser diode operating continuously at a wavelength of 375 nm to stimulate the samples. An objective microscope facilitated precise excitation focusing and simultaneous signal acquisition. Following this, the collected PL signal was meticulously filtered using a 405-nm upper-pass filter and acquired using a high-sensitivity visible-light spectrometer with the optical fiber. Transitioning to TRPL measurements, the laser diode's operation mode was adjusted, switching from continuous-wave (CW) mode to pulse operation with a 10 MHz repetition rate. The emitted PL signal, filtered with a specific long-pass filter with a cut-on wavelength of 405 nm, was directed towards a single-photon avalanche photodiode. The temporal behavior of this signal was thoroughly scrutinized using time-correlated single-photon counting electronics with an IRF of 0.3 ns[11]. For temperature-dependent PL and TRPL measurements, a Linkam HFS600E-PB4 cryostat with a liquid nitrogen cooling system was used. This setup enabled us to conduct measurements ranging from 80 K to 370 K with increments of 10 K.

**XL and TL.** The experimental methodology involved a comprehensive setup meticulously designed to accommodate the XL and TL assessments. This integrated configuration incorporated essential components, including an Inel XRG3500 X-ray generator operating at 45 kV / 10 mA with a copper anode tube, an Acton Research Corp. SpectraPro-500i monochromator, a Hamamatsu R928 photomultiplier tube (PMT), and an APD Cryogenic Inc. closed-cycle helium cooler. After the temperature was reduced to 10 K, the crystals were exposed to X-rays for 10 minutes. Subsequently, the TL glow curves were captured using a gradual heating rate of 0.14 K/s over a temperature range spanning from 10 to 350 K. Following this, XL spectra were acquired at discrete temperatures, beginning at 350

K and descending in 10-K increments to 10 K. This systematic sequence of measurements was deliberately chosen to mitigate potential influences arising from the thermal release of charge carriers, ensuring a comprehensive assessment of the emission output.

**Time-resolved XL.** A time-correlated single photon counting method was utilized to obtain the RT time-resolved X-ray decay curve measurement with a Start:Stop ratio of approximately 5000:1. X-ray pulses were generated using a PicoQuant LDHP-C-440M pulsed diode laser derived from a Hamamatsu N5084 X-ray tube stimulated by light with a high voltage set to 35 keV. Laser activation was facilitated by a PicoQuant laser driver, with its reference output serving as the start signal, synchronized with an Ortec 567 time-to-amplitude converter (TAC). An Ortec 462 time calibrator was used to ensure precision in timing and calibrate the bin width. The emitted photons were captured and analyzed using a digital PicoQuant analyzer. The total IRF of the system is similar to the TRPL of 0.3 ns[11]. The alignment of the sample position took place within a closed-cycle helium cryostat operating at pressures below $10^{-4}$ mbar, maintaining optimal experimental conditions.

**X-ray imaging.** In the X-ray imaging setup, 500-$\mu$m thick perovskite films on glass (see Figure S12, Supporting Information) were positioned in front of an LD Didactic 554 800 X-ray apparatus with a Mo source operating at 17.5 keV, 1 mA current and 35 keV voltage. A Type-18A line pattern card, with a lead thickness of 125 mm, was placed between the X-ray source and the film to minimize light scattering by positioning the pattern and the film as close as possible. The samples should be thinner than those in Figs. 1f and 1g for best resolution performance[47]. To capture the film's scintillation, an Allied Vision Mako U 130B camera with a 1.4-second exposure time and a conversion lens of 8 mm focal length was used. Image analysis involved calculating the edge spread function and its first derivative, the line spread function. The modulation transfer function was then determined by computing the modulus of the Fourier transform of the line spread function.

**PHS.** In pulse height spectra, 59.5 keV $\gamma$-excitation from $^{241}$Am (400 kBq) source was used. The samples were glued to the Hamamatsu R633 photomultiplier and covered with several layers of Teflon tape to form a protective layer. The signal was processed with the Nucliflare Digital Pulse Processor for $\gamma$-ray spectroscopy.

**Statistical Analysis. PL and RL**, the PL and RL data are presented as measured without any corrections or normalization. **TRPL and TRXL**, before fitting, we performed background subtraction and normalization of decay curves. **QE**, QE was determined using a commercially available setup from Edinburgh Instruments F980 with the corresponding software. **PHS**, PHS spectra were fitted "as measured" without further data treatments. Normalization was performed after fitting. A simple Gaussian function was employed to accommodate the full energy peak and to estimate the light yield. All fittings were performed with the Origin 2021 Pro software, provided that error values for each parameter were derived using the same software.

## ACKNOWLEDGMENTS

**All authors** acknowledge research funds from the National Science Center, Poland, under grant OPUS-24 no. 2022/47/B/ST5/01966. **M.M.** acknowledges the funds from the National Science Center, Poland, under grant Miniatura-8 no. 2024/08/X/ST5/00980 for consumable purchase. **F.M. and C.D.** acknowledge the financial support from the Ministry of Education, Singapore, under its AcRF Tier 2 grant (MOE-T2EP50121-0012). **F.A.A.N.** acknowledges funding from the Indonesian Endowment Fund for Education (LPDP) on behalf of the Indonesia Ministry of Education, Culture, Research and Technology, which is managed by Universitas Indonesia under INSPIRASI Program (Grant No PRJ-61/LPDP/2022 and 612/E1/KS.06.02/2022). **F.A.A.N.** also acknowledges Wildan P. Tresna for his help in the initial phase of the FDTD calculations. **W.D.** acknowledges Mohanad Eid for his assistance in the radio- and thermoluminescence measurements. **M.M.** acknowledges his leave status from Nicolaus Copernicus University in Torun, Torun 87-100, Poland.

## V. AUTHOR CONTRIBUTIONS STATEMENT

**M.M.**: Methodology, Formal analysis, Investigation, Conceptualization, Writing-original draft, Writing-review & editing, **W.Y., D.K., F.M., S.M., Y.T.A., W.Z., M.E.W, K.J.D, N., C.D., J.C., C.D.**: Formal analysis, Investigation, Writing-review & editing; **W.D., F.A.A.N., L.J.W.** Supervision, Resources, Writing-review; **M.D.B.** Conceptualization, Supervision, Resources, Writing-review & editing, Project administration, Funding acquisition.

## VI. ADDITIONAL INFORMATION

**Supporting Information** Supporting Information is available free of charge on the
**Competing interests** The authors declare that they have no competing interests.

# Supporting Information for Scaling Up Purcell-Enhanced Self-Assembled Nanoplasmonic Perovskite Scintillators into the Bulk Regime

Michal Makowski,[1, a)] Wenzheng Ye,[2, 3] Dominik Kowal,[1] Francesco Maddalena,[2, 3] Somnath Mahato,[1] Yudhistira Tirtayasri Amrillah,[4] Weronika Zajac,[1, 5] Marcin Eugeniusz Witkowski,[6] Konrad Jacek Drozdowski,[6] Nathaniel,[4] Cuong Dang,[2, 3] Joanna Cybinska,[1, 5] Winicjusz Drozdowski,[6] Ferry Anggoro Ardy Nugroho,[4, 7] Christophe Dujardin,[8, 9] Liang Jie Wong,[2, 3, b)] and Muhammad Danang Birowosuto[1, c)]

[1)] *Lukasiewicz Research Network - PORT Polish Center for Technology Development, Wroclaw, 54-066, Poland*

[2)] *CINTRA (CNRS-International-NTU-THALES Research Alliance), IRL 3288 Research Techno Plaza, 50 Nanyang Drive, Border X Block, Level 6, Singapore 637553, Singapore*

[3)] *School of Electrical and Electronic Engineering, Nanyang Technological University, Singapore 639798, Singapore*

[4)] *Department of Physics, Faculty of Mathematics and Natural Sciences, Universitas Indonesia, Depok 16424, Indonesia*

[5)] *Faculty of Chemistry, University of Wroclaw, Wroclaw, 50-383, Poland*

[6)] *Institute of Physics, Faculty of Physics, Astronomy, and Informatics, Nicolaus Copernicus University in Torun, Torun, 87-100, Poland*

[7)] *Institute for Advanced Sustainable Materials Research and Technology (INA-SMART), Faculty of Mathematics and Natural Sciences, Universitas Indonesia, Depok 16424, Indonesia*

[8)] *Universite Claude Bernard Lyon 1, Institut Lumiere Matiere UMR 5306 CNRS, 10 rue Ada Byron, Villeurbanne, 69622, France*

[9)] *Institut Universitaire de France (IUF), 1 rue Descartes, Paris Cedex 05, 75231, France*

(Dated: 13 May 2025)

[a] Electronic mail: michal.makowski@port.lukasiewicz.gov.pl
[b] Electronic mail: liangjie.wong@ntu.edu.sg
[c] Electronic mail: muhammad.birowosuto@port.lukasiewicz.gov.pl

# LIST OF FIGURES

## LIST OF TABLES

# LIGHT-MATTER INTERACTIONS IN NANOPLASMONIC SCINTILLATION

High-energy particles, particularly X-rays, interact with scintillators, converting their energy into visible light. The high-energy electrons generated by these particles lose energy through excitation, forming multiple secondary electrons and holes within the scintillation material, each with unique spatial positions and orientations. These electrons and holes subsequently form electron-hole pairs or dipole emitters. These pairs then undergo relaxation, either radiatively emitting visible photons or nonradiatively dissipating energy as heat.

In the case of a bare scintillator, the emission characteristics, such as emission rate, emission power, and emission direction, are solely determined by the intrinsic properties of the scintillation material, including its atomic structure. The dipole emission is mainly coupled into the radiative mode with an emission rate of $\Gamma_{\text{bare}}$, representing the number of photons produced by the dipole emitter. The photon flux reaching the photodetector is given by $\Gamma_{\text{bare}}$ multiplied by the energy transmission coefficient $T_{\text{bare}}$, which quantifies the efficiency of visible photon transmission from the dipole emitter to the detector. Thus, the emission power (or visible photon flux) is proportional to the emission rate and energy transmission coefficient:

$$\frac{dN_{\text{bare}}}{dt} = \Gamma_{\text{bare}} T_{\text{bare}} = \frac{P_{\text{bare}}}{\hbar\bar{\omega}} \tag{S1}$$

where $N$, $t$, $P$, and $\hbar\bar{\omega}$ represent the detected photon number, time, detected emission power, and average optical photon energy, respectively. Then, with the introduction of a nanophotonic design, especially a nanoplasmonic structure (as studied here), the scenario changes significantly. Nanoplasmonic designs, particularly those with structures on length scales comparable to visible wavelengths, enhance emission properties in two ways. First, they modify the intrinsic emission rate of the dipole emitters within the scintillators through the well-known Purcell effect, represented as $\Gamma = Fp\Gamma_{\text{bare}}$, where $Fp$ is the Purcell factor. The Purcell factor quantifies the enhancement of the spontaneous emission rate of a quantum emitter, such as $CsPbBr_3$ NC, in the presence of electromagnetic modes. This enhancement is directly proportional to the local density of states (LDOS) in the position of the emitter, that is, $Fp \propto \text{LDOS}$. The local density of states can be calculated as:

$$\text{LDOS}(\mathbf{r},\omega) = \frac{6\omega}{\pi c^2} [\mathbf{n}_\mu \cdot \text{Im}\{\mathbb{G}(\mathbf{r},\mathbf{r},\omega)\} \cdot \mathbf{n}_\mu] \tag{S2}$$

where $\mathbf{r}$ is the location of the dipole, $\boldsymbol{\mu}$ is the dipole moment, $c$ is the speed of light in a vacuum, and $\mathbb{G}(\mathbf{r},\mathbf{r'},\omega)$ is the Green's function at the dipole's location.

Second, the nanoplasmonic structure influences the transmission coefficient of the internal photons $T$, modulating how they enter the external environment for photodetector capture. This transmission coefficient is given by $T = \eta T_{\text{bare}}$. The detected photon flux in the presence of a nanophotonic design is thus:

$$\frac{dN}{dt} = \Gamma T = \eta Fp \Gamma_{\text{bare}} T_{\text{bare}} = \frac{P}{\hbar\bar{\omega}} \tag{S3}$$

As a result, the photodetector photon flux is enhanced by a factor of $\eta Fp$. This work explores the Purcell factor, $Fp$, for a dipole emitter near both spheroid and cuboid nanoparticles (NPs). For further derivations of $Fp$, we initially presume that the quantum efficiency ($QE$) is unity, but the effect of $QE$ to $Fp$ will be explained later.

**Purcell Factor for a Dipole near a spheroid NP**

For a single spheroid NP, the Purcell factor for a dipole close to NP[1,2] includes two components based on dipole polarization. The perpendicular component, $Fp_\perp^{sphere}$, is given by:

$$Fp_\perp^{sphere} = \frac{3}{2}\sum_{n=1}^{\infty} n(n+1)(2n+1)\left|\frac{j_n(y_{med}) + b_n h_n^{(1)}(y_{med})}{y_{med}^2}\right|^2 \tag{S4}$$

Similarly, the parallel component, $Fp_\parallel^{sphere}$, is expressed as:

$$Fp_\parallel^{sphere} = \frac{3}{4}\sum_{n=1}^{\infty}(2n+1)\left\{\left|j_n(y_{med}) + a_n h_n^{(1)}(y_{med})\right|^2 + \frac{\left|[y_{med}j_n(y_{med})]' + b_n\left[y_{med}h_n^{(1)}(y_{med})\right]'\right|^2}{y_{med}^2}\right\} \tag{S5}$$

Here, $y_{med} = k_{med}r'$, where $k_{med} = k_0 n_2$, and $k_0 = \omega/c$. $r$, $r'$, $\omega$, $c$, $\perp$, and $\parallel$ denote the position of the electric field, the position of the dipole emitter (a CsPbBr$_3$ NC) within a medium (PDMS) at approximately 5 nm from the NP, the emission angular frequency, the speed of light, the radial and tangential dipoles, respectively. $j_n$ and $h_n^{(1)}$ are spherical Bessel functions.

Since $Fp$ is related to maximum enhancement, we define $Fp^{sphere} = Fp_\perp^{sphere}$ for a dipole emitter with perpendicular orientation to the surface. In the isotropic case, the Purcell factor becomes $Fp^{sphere} = Fp_{iso}^{sphere} = \frac{2}{3}Fp_\parallel^{sphere} + \frac{1}{3}Fp_\perp^{sphere}$.

For analytical calculations of the emitters inside the medium with a refractive index of 1.5 and close to a spheroid NP with a diameter of 100 nm, we performed several distance parameters and dipole orientations to optimize our experimental conditions. For distances, since CsPbBr$_3$ NCs have diameters of about 7 nm, we varied the distances from 5 to 12 nm. $Fp_\parallel^{sphere}$, $Fp_\perp^{sphere}$, and $Fp_{iso}^{sphere}$ for the emitters located at 5 nm above the spheroid NP surfaces are 0.05, 8.06. and 2.72, respectively. Meanwhile, those for emitters located at 12 nm are 0.22, 6.73, and 2.39, respectively.

**Purcell factor for a dipole near a cuboid NP**

To approximate the Purcell factor for a dipole near a cuboid NP, we can utilize findings from prior studies[3,4]. However, these previous solutions focus primarily on configurations that involve a substrate. Another possible approach considers plasmon resonances as an eigenvalue problem[5], allowing us to approximate the Purcell factor for a cuboid NP via radiation corrections to electrostatic models. However, this method is valid only when the incident radiation wavelength is significantly larger than the NP dimensions.

In our approach, we return to a simpler approximation, analyzing the case where the dipole is near a cuboid NP surface, positioned away from the edges and the corners. For a dipole orientation parallel to the surface, the Purcell factor is given by

$$Fp_\parallel^{cube}(d,\omega) = 1 + \frac{3}{4k_0}\text{Re}\left\{\int_0^\infty dk_s \left[\frac{k_s}{k_\perp}\left(R^{TM} - \frac{k_\perp^2}{k_0^2}R^{TE}\right)e^{2ik_\perp d}\right]\right\}, \quad (S6)$$

where $k_0 = \omega/c$, $k_s$ is the in-plane wavenumber, $k_\perp$ is the normal wavenumber, and $R^{TM}$ and $R^{TE}$ are the Fresnel reflection coefficients for TM and TE polarized light, respectively. Here, $\omega$ is the emission angular frequency, $c$ is the speed of light, and $d$ is the distance between the dipole and the surface. The Fresnel reflection coefficients can be expressed as

$$R^{TE} = \frac{k_{0,z} - k_{metal,z}}{k_{0,z} + k_{metal,z}}, \quad R^{TM} = \frac{k_{0,z}\epsilon_{metal,y} - k_{metal,z}}{k_{0,z}\epsilon_{metal,y} + k_{metal,z}}, \quad (S7)$$

where

$$k_{0,z} = \sqrt{k_0^2 - k_s^2}, \quad k_{metal,z} = \sqrt{k_0^2\epsilon_{metal,r} - k_s^2}, \quad (S8)$$

and $\epsilon_{metal,r}$ is the relative permittivity of the metal. For a dipole orientation perpendicular to the surface, the Purcell factor becomes

$$Fp_\perp^{cube}(d,\omega) = 1 + \frac{3}{2k_0}\text{Re}\left\{\int_0^\infty dk_s \left[\frac{k_s^3}{k_\perp k_0^2}R^{TE}e^{2ik_\perp d}\right]\right\}. \quad (S9)$$

For an isotropic case, the Purcell factor $Fp^{cube}$ is given by $Fp_{iso}^{cube} = \frac{2}{3}Fp_\parallel^{cube} + \frac{1}{3}Fp_\perp^{cube}$.

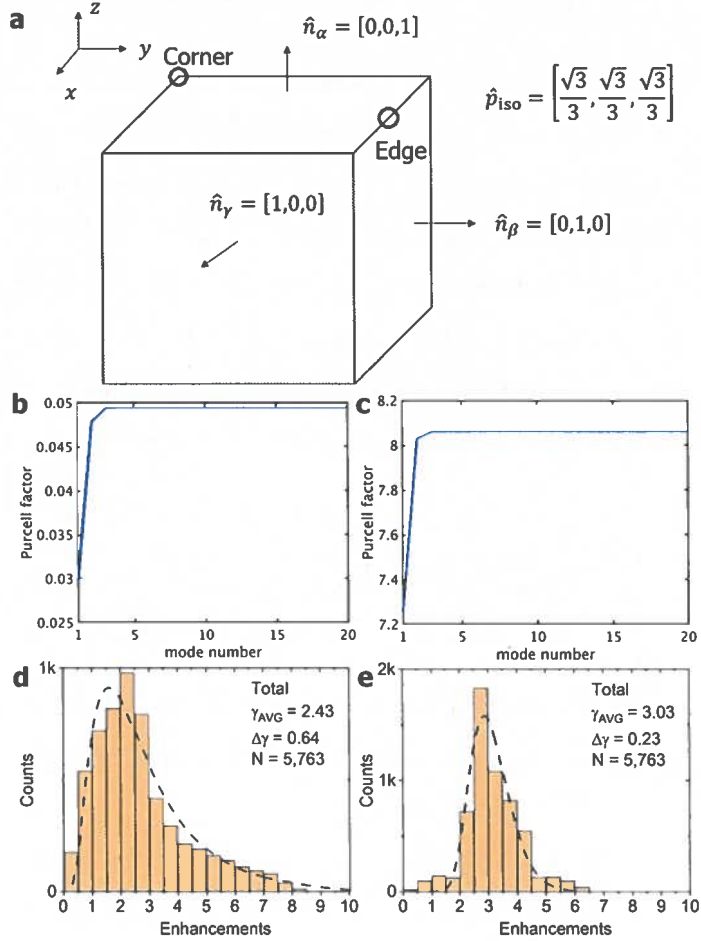

FIG. S1. **Representations of dipole orientation in the isotropic case and analytical calculation results. a,** Dipole orientations with their cuboid NP surfaces. The circular areas define the edge and the corner. **b, c,** Results of convergence for the lowers (**b**) and the highest (**c**) Purcell factors obtained with analytical calculation. **d, e,** Histograms of enhancement factors for Ag SNPs (**d**) and Ag CNPs (**e**) samples after Monte Carlo simulations over 5,763 emitters.

If the dipole is placed near the edge of a nanoplasmonic cube, the interaction with multiple surfaces alters the effective orientation. Here, we define the normal vectors of the metal surfaces 1 and 2 as $\hat{n}_1$ and $\hat{n}_2$, respectively, with the unit vector of the dipole orientation $\hat{p}$. The distances from the dipole to surfaces 1 and 2 are $d_1$ and $d_2$, respectively. The isotropic Purcell factor $Fp^{cube,edge}$ for this configuration can be approximated by

$$Fp^{cube,edge}(\mathbf{r},\omega,\hat{p}) = 1 + \left\{ [Fp_\parallel^{cube}(d_1,\omega) - 1]|\hat{p} \times \hat{n}_1|^2 + [Fp_\parallel^{cube}(d_2,\omega) - 1]|\hat{p} \times \hat{n}_2|^2 \right.$$
$$\left. + [Fp_\perp^{cube}(d_1,\omega) - 1]|\hat{p} \cdot \hat{n}_1|^2 + [Fp_\perp^{cube}(d_2,\omega) - 1]|\hat{p} \cdot \hat{n}_2|^2 \right\}. \quad (S10)$$

For dipoles with isotropic orientations, we can use $\hat{p}_{iso} = [\sqrt{3}/3, \sqrt{3}/3, \sqrt{3}/3]$, $\hat{n}_\alpha = [0.0,1]$, $\hat{n}_\beta = [0.1,0]$, and $\hat{n}_\gamma = [1.0,0]$ for Eq. S10 by noting that surfaces 1 and 2 are two of the surfaces $(\alpha,\beta,\gamma)$, depending on the dipole position; see Supporting Fig. S1. The definition of the isotropic Purcell factor then becomes $Fp_{iso}^{cube,edge}(\mathbf{r},\omega) = Fp^{cube,edge}(\mathbf{r},\omega,\hat{p}_{iso})$.

For a dipole placed in the corner of a nanoplasmonic cube, we extend the edge-based approach to include interactions with three surfaces. Here, the normal vectors of surfaces 1, 2, and 3 are $\hat{n}_1$, $\hat{n}_2$, and $\hat{n}_3$, respectively, and the distances from the dipole to these surfaces are $d_1$, $d_2$, and $d_3$. The isotropic Purcell factor $Fp^{cube,corner}$ is given by

$$Fp^{cube,corner}(\mathbf{r},\omega,\hat{p}) = 1 + \left\{ [Fp_\parallel^{cube}(d_1,\omega) - 1]|\hat{p} \times \hat{n}_1|^2 + [Fp_\parallel^{cube}(d_2,\omega) - 1]|\hat{p} \times \hat{n}_2|^2 \right.$$
$$+ [Fp_\parallel^{cube}(d_3,\omega) - 1]|\hat{p} \times \hat{n}_3|^2 + [Fp_\perp^{cube}(d_1,\omega) - 1]|\hat{p} \cdot \hat{n}_1|^2$$
$$\left. + [Fp_\perp^{cube}(d_2,\omega) - 1]|\hat{p} \cdot \hat{n}_2|^2 + [Fp_\perp^{cube}(d_3,\omega) - 1]|\hat{p} \cdot \hat{n}_3|^2 \right\}. \quad (S11)$$

For the dipoles at the corner, using the same analogy as the dipoles at the edge, we can use $\hat{p}_{iso} = [\sqrt{3}/3, \sqrt{3}/3, \sqrt{3}/3]$, $\hat{n}_1 = [0.0,1]$, $\hat{n}_2 = [0.1,0]$, and $\hat{n}_3 = [1.0,0]$. For the Purcell factor, $Fp_{iso}^{cube,corner}(\mathbf{r},\omega) = Fp^{cube,corner}(\mathbf{r},\omega,\hat{p}_{iso})$.

Similarly to a spheroid NP, we performed analytical calculations of the emitters inside the medium with a refractive index of 1.5 with 5- and 12-nm distances from cuboid NP surfaces with a side length of 100 nm. However, since there are three different cuboid surfaces in Fig. S1a: flat surfaces, edges, and corners, we calculated them separately. First, for flat surfaces, $Fp_\parallel^{cube}$, $Fp_\perp^{cube}$, and $Fp_{iso}^{cube}$ for emitters located 5 nm above the NP are 3.54, 0.39, and 2.49, respectively, while those located 12 nm above the NP are 3.01, 0.30, and 2.11, respectively. Second, for the edges, $Fp_\parallel^{cube,edge}$, $Fp_\perp^{cube,edge}$, and $Fp_{iso}^{cube,edge}$ for emitters located 5 nm above the NP are 6.08, 2.93, and 3.98, respectively, while those located 12 nm above the NP are 5.03, 2.32, and 3.22, respectively. Finally, for the corner, $Fp_\parallel^{cube,corner}$, $Fp_\perp^{cube,corner}$, and $Fp_{iso}^{cube,corner}$ for emitters located 5 nm above the NP have the same values of 5.47, while those located 12 nm above the NP also have the same values of 4.33. All the calculated enhancement values, including those calculated for the spheroid NPs, are convergent, as shown in Figs. S1b and S1c.

**From point dipole emitters to finite CsPbBr$_3$ NCs**

Before performing the complete statistical analysis for average enhancements of Ag SNPs and CNPs, we evaluated how we treat CsPbBr$_3$ NCs in our analytical calculations. The differences from the previous point dipole emitter calculations are the finite diameter $d$ and the broadband emission frequency $\omega$ of CsPbBr$_3$ NCs. This approach is very similar to our previous method in the calculation of the average emission rates for single fluorescent spheres of 25 nm[6]. Our analytical calculations consider our CsPbBr$_3$ NCs as a single 7-nm-diameter sphere of many dipole emitters separated by 0.5-nm distances. If the positions are at 0 and 7 nm from the 5-nm PVP and Ag NP surface, they are zero dipole emitters, while the probability function for dipole emitters to be inside the single CsPbBr$_3$ NC for each $t$ distance from the center $C_{prob}$ is given by

$$C_{prob} = \frac{3}{4} \frac{\left(\left(\frac{d}{2}\right)^2 - t^2\right) \Delta h}{\left(\frac{d}{2}\right)^3} \tag{S12}$$

, whereas $d$ and $\Delta h$ of 7 and 1 nm are the cross-section diameter and the atomic-like thickness of single CsPbBr$_3$ NC, respectively. Finally, Purcell factor of single CsPbBr$_3$ NC for the specific emission frequency $f(\omega)$ can be written as

$$f(\omega) = \int_{-\frac{d}{2}}^{\frac{d}{2}} Fp\left(l + \frac{d}{2} - t, \omega\right) C_{prob}(t) dt \tag{S13}$$

, whereas $l$ of 5 nm is the thickness of the PVP layer surrounding Ag SNPs and CNPs and $Fp\left(l + \frac{d}{2} - t, \omega\right)$ is the Purcell factor at each point inside CsPbBr$_3$ NCs as functions of $t$ and $\omega$.

Because we have two dipole polarizations, parallel and perpendicular, and the emitters are randomly located in both Ag SNPs and CNPs, the calculation is more complicated than Eq. S13. We should calculate $Fp$ for each possible dipole orientation, while the $x$-$y$-$z$ coordinates are chosen to match the direction of the main axes in the system[7]. Then, $Fp$ can be simply defined by $Fp_{max}$, $Fp_{med}$, and $Fp_{min}$ with the following expression

$$Fp = Fp_{min}x^2 + Fp_{med}y^2 + Fp_{max}z^2 \tag{S14}$$

where $x$-$y$-$z$ coordinates lie on a sphere $x^2 + y^2 + z^2 = 1$ because they are the components of the dipole orientation vector $e_d$. The $x$-$y$-$z$ coordinates can be conveniently transformed into the spherical coordinate system. Since our $Fp$ calculated for distances between 5 and 12 nm, for Ag SNP, $Fp_{min} = Fp_{med} = Fp_{\parallel}^{sphere}$ and $Fp_{max} = Fp_{\perp}^{sphere}$. This situation changes for Ag CNP which the dipoles are located on flat surfaces and edges where $Fp_{min} = Fp_{med} = Fp_{\perp}^{cube}$ or $Fp_{\perp}^{cube,edge}$ and $Fp_{max} = Fp_{\parallel}^{cube}$ or $Fp_{\parallel}^{cube,edge}$. Finally, for the dipoles located on the corners of Ag CNP, all $Fp^{cube,corner}$ values are the same.

To calculate the distribution of $f(\omega)$ that is expected from the experiments with many CsPbBr$_3$ NCs coupled to a single Ag SNP or CNP, a Monte Carlo method was used[6,8]. Figs. 1d and 1e show the distribution of the enhancements for 5,783 single CsPbBr$_3$ NCs in the vicinity of Ag SNPs and CNPs, respectively. For this number of emitters, the error in % is expected to be small since the previous error still falls between 5 and 11% for 10,000 and 1,000 emitters, respectively[8].

## ABSORPTION LENGTH AND SCATTERING SPECTRA

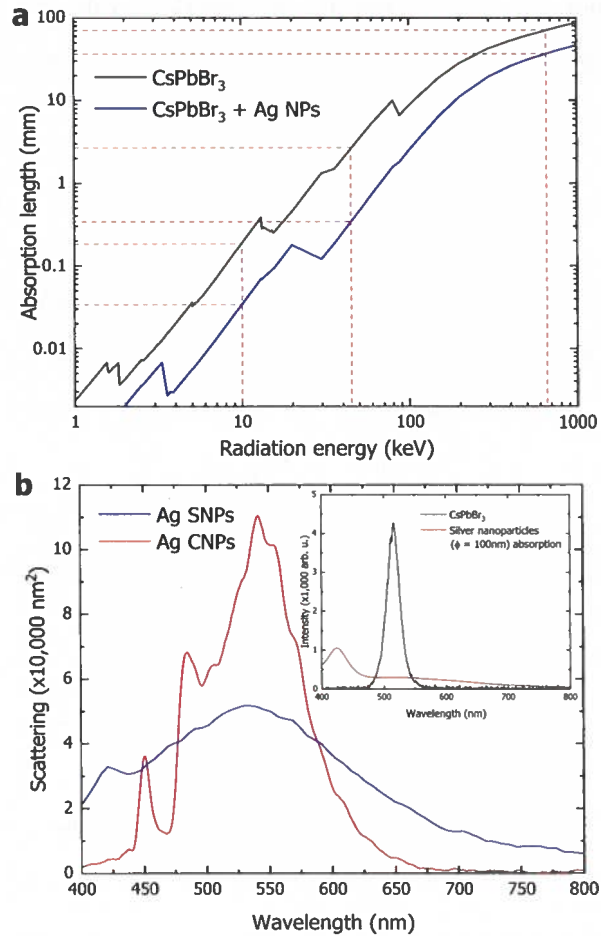

FIG. S2. **Theoretical absorption length and the additional optical characterizations of the samples. a**, Absorption length of CsPbBr3 NCs in PDMS (black line), and Ag NPs-doped sample (blue line). The dashed lines present the critical absorption lengths for 10, 50, and 662 keV, respectively. **b**, The scattering spectra of Ag SNPs and Ag CNPs compared. Inset presents PL spectra of pure CsPbBr3 NCs in PDMS (black line) and absorption spectra of Ag NPs with a diameter of 100nm (red curve).

A reduction in absorption length is visible in Fig. S2b. Three relevant X-ray energies of 10, 45, and 662 keV were selected for evaluation. The absorption lengths were reduced: 0.18

$\rightarrow$ 0.03 mm, 2.68 $\rightarrow$ 0.34 mm, and 70.32 $\rightarrow$ 36.35 mm for 10, 45, and 66 keV, respectively.

For scattering spectra, Ag NP measurements were performed in home-built microscope settings with a xenon light source (Thorlabs) focused through a 100× objective lens (NA = 0.95) in dark field geometry[9]. The scattered light was then collected by the same lens and directed into the hyperspectral system (Cytoviva) for spectral measurement of the individual particles.

## FDTD CALCULATIONS

Absolute decay rates, such as $\Gamma_{rad}$, $\Gamma_{rad^o}$, $\Gamma_{nrad}$, cannot be calculated with FDTD. However, the quantum mechanical decay rate in an inhomogeneous environment $\Gamma_{dip}$ and the classical power radiated by the dipole in the same environment $P_{dip}$ are related by:

$$\frac{\Gamma_{dip}}{\Gamma^o_{rad}} = \frac{P_{dip}}{P^o_{rad}} \tag{S15}$$

Where $\Gamma^o_{rad}$ and $P^o_{rad}$ are the decay rate and radiated power of the dipole in a homogeneous environment.

The Purcell factor is the emission rate enhancement shown as

$$F_p = \frac{\Gamma_{dip}}{\Gamma^o_{rad}} \tag{S16}$$

Therefore, the Purcell factor can be calculated using a simple built-in source-power function in Lumerical FDTD. Quantum efficiency $QE$ can be calculated as the relation between the decay rates or the power as

$$\frac{QE}{QE^o} = \frac{\Gamma_{rad}}{\Gamma_{rad} + \Gamma_{nrad}} = \frac{P_{rad}}{P_{rad} + P_{nrad}} \tag{S17}$$

where $\Gamma_{nrad}$ and $P_{nrad}$ is an additional nonradiative decay rate of the transition in the emitter and the power absorbed by the antenna only. Taking the emitter to have a high intrinsic $QE$, therefore $QE^o$ is equal to 100 % and $\Gamma^o_{nr} = 0$. So that

$$QE = \frac{\Gamma_{rad}}{\Gamma_{rad} + \Gamma_{nrad}} = \frac{P_{rad}}{P_{rad} + P_{nrad}} \tag{S18}$$

This formulation corresponds directly to Eq. (3) in the main text, where the Purcell factor is computed the basis of energy density integral.

TABLE S1. **Comparison of room temperature time-resolved photo- and X-ray luminescence fitting parameters.** Where $\tau_i$ (for i = 1,2 and 3), $\bar{\tau}$, $\Gamma$ and $\Gamma/\Gamma_0$ corresponds to i-th decay component, mean decay time, decay rate and relative decay rate, respectively. Percentages correspond to the i-th decay contribution.

|  | PL | | | XL | | |
|---|---|---|---|---|---|---|
|  | $CsPbBr_3$ | $CsPbBr_3$ +Ag SNPs | $CsPbBr_3$ + Ag CNPs | $CsPbBr_3$ | $CsPbBr_3$ +Ag SNPs | $CsPbBr_3$ + Ag CNPs |
| $\tau_1$ (ns) | 0.99 ± 0.03 (42.6%) | 0.71 ± 0.05 (76.0%) | 0.61 ± 0.04 (89.1%) | 0.71 ± 0.01 (22.1%) | 0.70 ± 0.03 (45.8%) | 0.67 ± 0.04 (52.1%) |
| $\tau_2$ (ns) | 6.28 ± 0.07 (57.4%) | 4.46 ± 0.12 (24.0%) | 3.71 ± 0.15 (10.1%) | 3.16 ± 0.04 (48.5%) | 3.05 ± 0.09 (38.0%) | 2.91 ± 0.07 (30.9%) |
| $\tau_3$ (ns) | - | - | - | 20.01 ± 0.31 (29.4%) | 16.62 ± 0.41 (16.2%) | 14.53 ± 0.33 (17.0%) |
| $\bar{\tau}$ (ns) | 4.02 ± 0.13 | 1.61 ± 0.08 | 0.94 ± 0.06 | 7.55 ± 0.12 | 4.18 ± 0.13 | 3.71 ± 0.09 |
| $\Gamma$ (1/ns) | 0.25 ± 0.01 | 0.62 ± 0.01 | 1.05 ± 0.02 | 0.13 ± 0.01 | 0.24 ± 0.01 | 0.27 ± 0.01 |
| $\Gamma/\Gamma_0$ | 1.00 ± 0.01 | 2.48 ± 0.17 | 4.20 ± 0.31 | 1.00 ± 0.01 | 1.84 ± 0.07 | 2.08 ± 0.06 |

# FDTD CALCULATION OF COUPLED CSPBBR$_3$ NCS ON AG NP SURFACES

An analysis of the overlap between energy-dispersive X-ray spectroscopy (EDS) images for Ag and Pb or Br elements was conducted using the Sørensen-Dice index (SDI)[10-13], following the conversion of the images to black-and-white scale. This analysis incorporated size approximations of Ag nanocubes and CsPbBr$_3$ NPs. The percentage of coupled CsPbBr$_3$ NPs with Ag was derived from the averages of 8 overlap mappings between Ag and the Pb or Br elements. Thus, the coupling percentage $C_{coup}$ can be calculated as[13]:

$$C_{coup} = \text{SDI} \times 100\% = \frac{2|A_{Ag} \cap A_{Pb,Br}|}{|A_{Ag}|+|A_{Pb,Br}|} \times 100\% \tag{S19}$$

where $|A_{Ag}|$ and $|A_{Pb,Br}|$ are the cardinalates between Ag atomic sets and Pb or Br atomic sets.

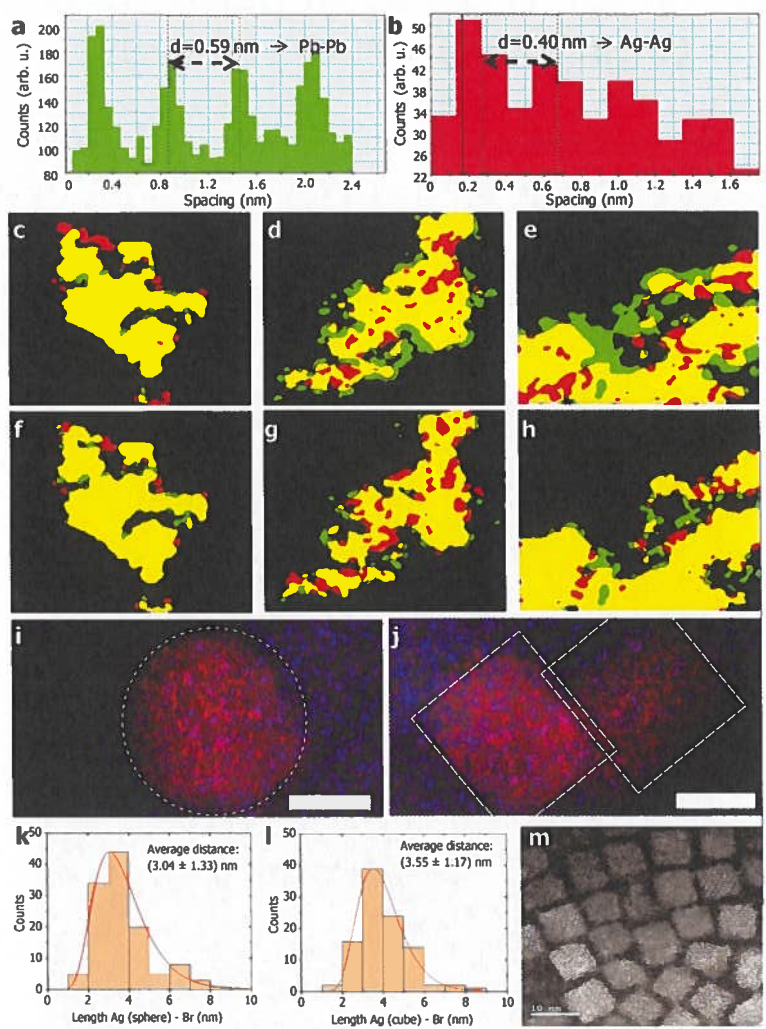

FIG. S3. **STEM HAADF an EDS images.** a,b, The atomic distances of Pb-Pb (a) and Ag-Ag (b), extracted from the STEM-HAADF images with Digital Micrograph software. c-h, EDS images of Ag-Br (c-e) and Ag-Pb (f-h) overlaps. The red, green, and yellow color corresponds to Ag, Pb, or Br atoms. The overlap area is derived from SDI. i-l, EDS images (i and j) and histograms of distances between Ag and Br atoms (k and l) for single SNP (i and k) and CNP (j and l) with many CsPbBr$_3$ NCs. The red and blue spots are the Ag and Cs atoms, respectively, while the dashed lines are added to improve the identification of the structures. The white scale bar is 50 nm. The histograms in k and l are calculated by the distances of the closest blue spots to the white dashed lines (within < 12 nm as similar to the size of the NCs and PVP layer). m, STEM HAADF image of CsPbBr$_3$ NCs. The white scale bar in m is 10 nm.

# QUANTUM EFFICIENCY, PL QUANTUM YIELD, AND ITS RELATIONSHIP TO THE PURCELL ENHANCEMENTS

PL quantum yield measurements were used to estimate $QE$. The measurements were performed using an Integration Sphere Assembly F-M01 (Edinburgh Instruments) with an FLS980 spectrofluorimeter. Xenon lamp was used as an excitation source, with a 375 nm wavelength selection. From several measurements, we found that the PLQY is about 55 ± 15 %. This is comparable to the previous report of 64 %[14].

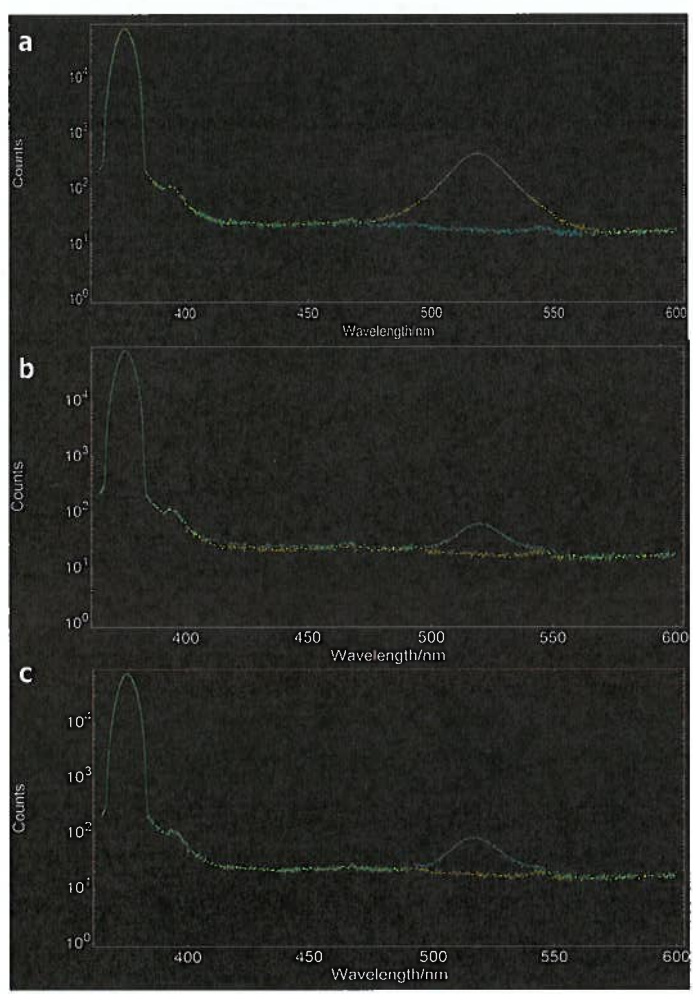

FIG. S4. **Measured Toluene reference and photoluminescence spectra for determination of quantum efficiencies.** a-c, Spectra collected from $CsPbBr_3$ (**a**), $CsPbBr_3$ + Ag SNPs (**b**), and $CsPbBr_3$ + Ag CNPs (**c**).

We obtain $QE$ of (55 ± 15) % from the measurements above. Such $QE$ value affects both the intensity $P$ and the emission rate $\Gamma$ enhancements. Both cases are related to the contributions of radiative $\Gamma_{rad}$ and non-radiative $\Gamma_{nrad}$ and they can be written as

$$QE = \frac{\Gamma_{rad}}{\Gamma_{tot}} = \frac{\Gamma_{rad}}{\Gamma_{rad} + \Gamma_{nrad}} \qquad (S20)$$

where $\Gamma_{tot}$ is the total decay rate. For bare CsPbBr$_3$ NC in PDMS system (denoted by superscript "$o$"), we can obtain the following relationship:

$$\frac{\Gamma_{nrad}^o}{\Gamma_{rad}^o} = \frac{1}{QE} - 1 \qquad (S21)$$

The enhancements of the total decay rate in the CsPbBr$_3$-AgNP scintillator (denoted by the superscript "$np$") system, $\eta_{dec}$, can be written as

$$\eta_{dec} = \frac{\Gamma_{tot}^{np}}{\Gamma_{tot}^o} = \frac{\Gamma_{rad}^{np} + \Gamma_{nrad}^{np}}{\Gamma_{rad}^o + \Gamma_{nrad}^o} \qquad (S22)$$

The enhancements of intensity, $\eta_{int}$, can be written as

$$\eta_{int} = \frac{QE^{np}}{QE^o} = \frac{\Gamma_{rad}^{np}/\Gamma_{rad}^o}{\eta_{dec}} \qquad (S23)$$

Then we can obtain the enhancement of radiative decay rate ($\Gamma_{rad}^{np}/\Gamma_{rad}^o$) as

$$\frac{\Gamma_{rad}^{np}}{\Gamma_{rad}^o} = \eta_{int} \times \eta_{dec} \qquad (S24)$$

By simplifying the expression of $\eta_{dec}$, we can obtain the enhancement of non-radiative decay rate, ($\Gamma_{nrad}^{np}/\Gamma_{nrad}^o$), as

$$\frac{\Gamma_{nrad}^{np}}{\Gamma_{nrad}^o} = \eta_{dec}\frac{(1/QE^o) - \eta_{int}}{(1/QE^o) - 1} \qquad (S25)$$

The coupling with a plasmon can lead to enhancements in both radiative decay and non-radiative decay. For a high $QE$ system ($QE \approx$ 100 %), which means the total decay rate is dominated by the radiative decay rate, i.e., $\Gamma_{tot} \approx \Gamma_{rad}$. We can observe the enhancement of radiative decay rate, $\Gamma_{rad}^{np}/\Gamma_{rad}^o$, from the time-resolved PL measurements. At the same time, it is difficult to observe the enhancement of non-radiative decay rate, $\Gamma_{nrad}^{np}/\Gamma_{nrad}^o$. The emission efficiency thus will not be greatly enhanced since $QE$ is already close to 100 %. For a low $QE$ system ($QE \ll 100\%$), i.e. $\Gamma_{tot} \approx \Gamma_{nrad}$. We can observe the enhancement of non-radiative decay rate from time-resolved PL measurements, while it is hard to observe the enhancement of radiative decay rate. In our case, the $QE$ of 55 ± 15 % enables us to observe both radiative and non-radiative decay enhancements when the dipole emission is coupled to the plasmon. The 5-nm polyvinylpyrrolidone (PVP) layer assures that $\Gamma_{rad}^{np}/\Gamma_{rad}^{o} \gg \Gamma_{nrad}^{np}/\Gamma_{nrad}^{o}$.

Regarding the Purcell factor ($F_p$) from Eqs. S4, S5, S6, and S9, we can obtain the corrected Purcell factor ($F_p^{corr}$) for our samples (with many CsPbBr$_3$ NCs coupled to many Ag NPs) following:

$$F_p^{corr} \approx QE \times C_{coup} \times F_p \tag{S26}$$

where in our case, $QE$ of $(55 \pm 15)\%$ and $C_{coup}$ of $(70 \pm 8)\%$ resulting $F_p^{corr} \approx (39 \pm 17)\% F_p$.

# TEMPERATURE-DEPENDENT PHOTO- AND X-RAY LUMINESCENCE MEASUREMENTS

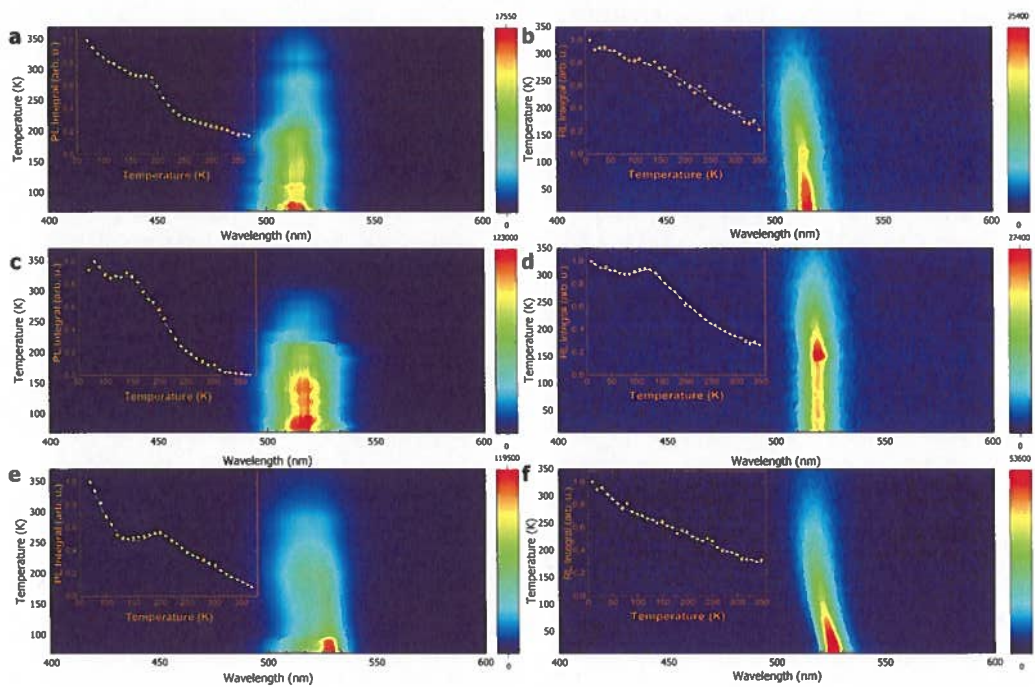

FIG. S5. **Temperature-dependent photo- and X-ray luminescence spectra of samples.** a-f, Spectral maps of photo- and X-ray luminescence of $CsPbBr_3$ (**a** and **b**, respectively), $CsPbBr_3$ + Ag SNPs (**c** and **d**, respectively) and $CsPbBr_3$ + Ag CNPs (**e** and **f**, respectively). Insets represent an integrated area under each luminescence alongside the NTQ fitting curve.

All samples, both pure and Ag NPs-doped, revealed a negative thermal quenching (NTQ) behavior, which was fit using Shibata's model[15]. All essential parameters and spectra are presented in Supporting Tables S2 and S3 and Supporting Fig. S5.

TABLE S2. **Analysis of temperature-dependent X-ray luminescence spectra.** X-ray luminescence negative thermal quenching model fitting parameters.

|  | $CsPbBr_3$ | $CsPbBr_3$ + Ag SNPs | $CsPbBr_3$ + Ag CNPs |
|---|---|---|---|
| $C_1$ | $1.1 \cdot 10^0$ | $6.6 \cdot 10^1$ | $9.2 \cdot 10^{-1}$ |
| $E_1^D$ (meV) | 19.9 | 27.9 | 8.09 |
| $C_2$ | $8.9 \cdot 10^1$ | $1.8 \cdot 10^3$ | $2.6 \cdot 10^1$ |
| $E_2^D$ (meV) | 113 | 87 | 84 |
| $D_1$ | - | $1.1 \cdot 10^2$ | - |
| $E_1^N$ (meV) | - | 33 | - |

TABLE S3. **Analysis of temperature-dependent photoluminescence spectra.** Photoluminescence negative thermal quenching model fitting parameters.

|  | $CsPbBr_3$ | $CsPbBr_3$ + Ag SNPs | $CsPbBr_3$ + Ag CNPs |
|---|---|---|---|
| $C_1$ | $1.9 \cdot 10^0$ | $1.8 \cdot 10^0$ | $2.5 \cdot 10^1$ |
| $E_1^D$ (meV) | 6.95 | 4.05 | 30.4 |
| $C_2$ | $6.8 \cdot 10^5$ | $2.6 \cdot 10^5$ | $1.6 \cdot 10^6$ |
| $E_2^D$ (meV) | 230 | 206 | 249 |
| $C_3$ | $2.6 \cdot 10^8$ | $7.6 \cdot 10^8$ | $8.9 \cdot 10^8$ |
| $E_3^D$ (meV) | 403 | 404 | 428 |
| $D_1$ | $5.1 \cdot 10^1$ | $4.2 \cdot 10^1$ | $5.4 \cdot 10^1$ |
| $E_1^N$ (meV) | 95.7 | 89.7 | 54.7 |

# LOW-TEMPERATURE THERMOLUMINESCENCE

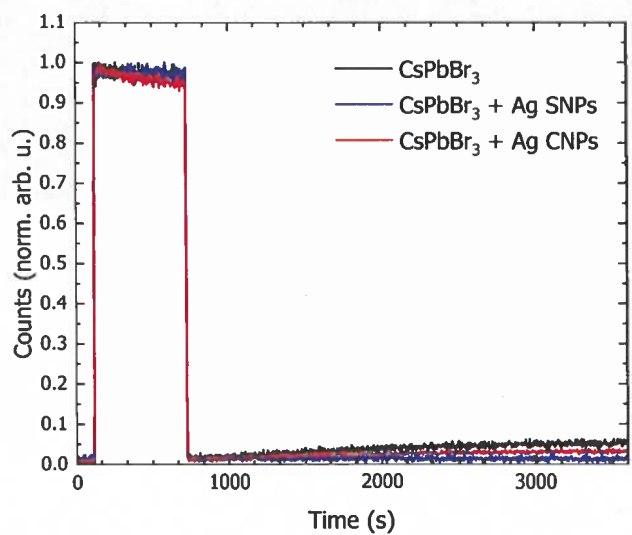

FIG. S6. **Thermoluminescence measurements.** Comparison of low-temperature thermoluminescence curves of pure $CsPbBr_3$, Ag SNP-doped, and Ag CNP-doped samples.

# X-RAY IMAGING

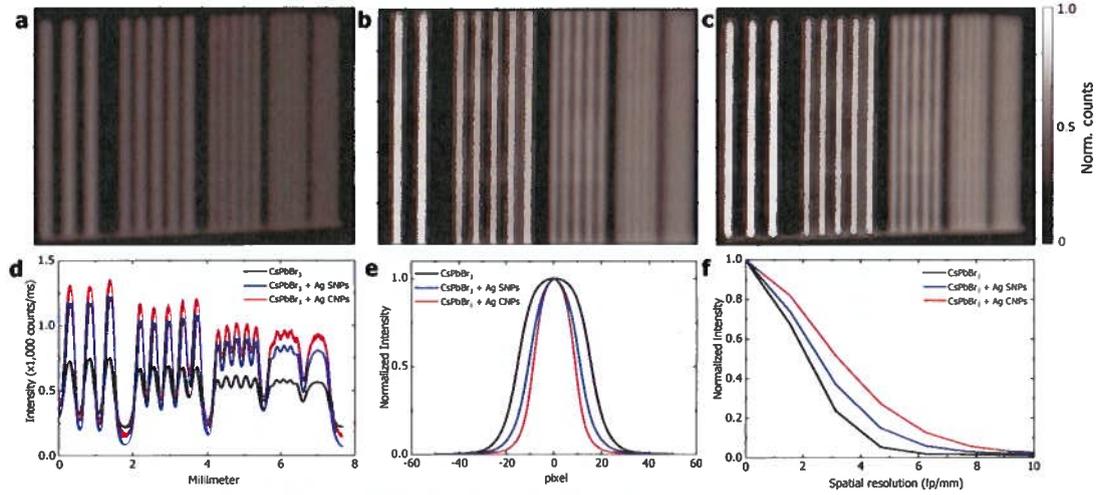

FIG. S7. **X-ray imaging using a resolution card a-c**, X-ray images employing resolution card type 18a of CsPbBr$_3$ (**a**), CsPbBr$_3$ + Ag SNPs (**b**), and CsPbBr$_3$ + Ag CNPs (**c**). **d**, Cross sections of card imaging. **e,f**, Line Spread Function (**e**), and Modulation Transfer Function (MTF) (**f**) of square X-ray imaging in Figs. 5a-5c.

TABLE S4. **Spatial resolution (lp/mm) at 0.2MTF.** The parameters derived from Figs. 5a-5c in the main manuscript after calculating edge and line spread functions with Fourier transform.

| Sample | Spatial resolution (lp/mm) at 0.2 MTF |
|---|---|
| CsPbBr$_3$ | 3.44 |
| CsPbBr$_3$ + Ag SNPs | 4.35 |
| CsPbBr$_3$ + Ag CNPs | 5.47 |

## PULSE HEIGHT SPECTRA

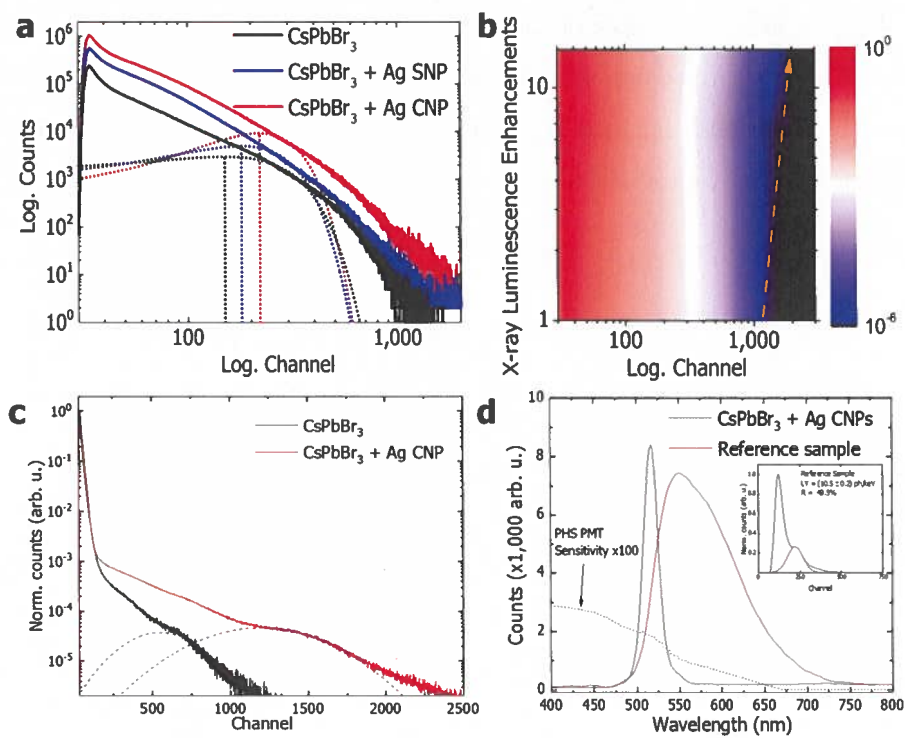

FIG. S8. **Pulse height spectra. a**, Pulse height spectra of various PDMS samples irradiated with $^{241}$Am, including CsPbBr$_3$ NCs, CsPbBr$_3$ NCs with Ag SNPs, and CsPbBr$_3$ NCs with Ag CNPs. Gaussian fits are overlaid on the pulse height spectra, with dashed lines indicating the fit curves and peak positions. **b**, A surface plot depicts the pulse height spectra as a function of theoretical enhancement, showcasing a pronounced long tail in the spectrum of CsPbBr$_3$ NCs with Ag CNPs, see orange arrow. **c**, Pulse height spectra with resolved photopeaks. The black curve shows PHS for pure CsPbBr$_3$ NCs sample, while the red curve correlates with PHS for CsPbBr$_3$ NCs with Ag CNPs, Gaussian fits are overlaid on the pulse height spectra, with dashed lines indicating the fit curves and peak positions. For the Ag CNP-doped sample, a light yield enhancement of (2.07 ± 0.39) was achieved. **d**, Comparison of RL spectra of CsPbBr$_3$ + Ag CNPs and a reference sample of Tb$_2$Y$_{0.5}$Al$_{0.5}$O$_{12}$ for light yield estimation. Inset present pulse height spectra of a reference sample, revealing a 10.5 ph/keV light yield.

The separation of the full-energy peak strongly depends on the alignment of NCs. When NCs are more randomly oriented and highly loaded into the PDMS[16], multiple electron scattering disrupts the appearance of the full-energy peak. By slowly mixing slightly underloaded NCs with PDMS to achieve better alignment, we obtained well-aligned NCs, enabling us to resolve the photopeak for pure and Ag CNP-doped samples. However, we still need to optimize the mixing for reproducibility.

# IMPACT OF PARTICLE AND EMITTER PARAMETERS ON THE ENHANCEMENT

To evaluate the influence of particle size on the Purcell enhancement, we conducted FDTD simulations for Ag NPs with varying sizes of 80 nm, 100 nm, and 120 nm. The results (shown in Fig. S9) indicate that the spatial distribution of the LDOS enhancement remains consistent across all sizes. Specifically, the enhancement remains isotropic for SNPs, whereas CNPs exhibit significant enhancement at the corners due to their sharp geometric features. Interestingly, the average LDOS enhancement exhibits a non-monotonic trend: it increases as the NP size grows from 80 nm to 100 nm but decreases for 120 nm NPs.

This behavior is attributed to the scattering properties of the NPs, which dictate emitter-NP coupling. The calculated scattering spectra (Fig. S10) reveal a redshift in the scattering peak for increasing particle size. Notably, the scattering intensity at the perovskite emission wavelength (520 nm) follows the same trend as the LDOS enhancement, confirming that the enhancement is governed by spectral overlap between NP scattering and emitter emission. Consequently, 100 nm NPs were selected due to their optimal spectral alignment with $CsPbBr_3$ NCs.

Another crucial factor affecting enhancement is the spatial distribution of NPs within the scintillator matrix. While full-scale computational modeling of randomly distributed NPs is prohibitively expensive, prior studies indicate that a single NP model can approximate a collection of non-interacting NPs. The justification lies in the far-field interference effects: in a randomly dispersed NP ensemble, far-field radiation contributions cancel out, allowing the system's optical properties to be represented by a single particle. However, this approximation holds only if interparticle coupling is negligible.

We estimated the required NP density for near-field interactions to assess potential coupling effects to become significant. Literature suggests particle coupling emerges when the average center-to-center distance is less than twice the NP's characteristic length (diameter for SNPs, side length for CNPs). For a 5×5×5 $mm^3$ scintillator matrix, reaching this coupling threshold would require at least $10^{13}$ NPs. In contrast, the experimental NP concentration corresponds to an average interparticle spacing of $\sim 1\mu m$, far exceeding the coupling limit (Fig. S10e). This confirms that the assumption of independent single-particle behavior remains valid in our experimental configuration.

Additionally, Monte Carlo simulations were performed to examine the effect of emitter density on enhancement. The number of coupled $CsPbBr_3$ NCs per NP varied from 1 to 200 for SNPs and CNPs, with 10.000 random spatial configurations per case. The results show that for SNPs, the LDOS enhancement remains nearly constant regardless of emitter count, consistent with their isotropic nature (Fig. S10f). In contrast, CNPs exhibit a gradual increase in enhancement with higher emitter density. This trend arises since increased emitter loading raises the probability of NCs occupying the high-enhancement regions near CNP corners, leading to a net increase in the average enhancement factor.

These findings underscore the significance of NP size and emitter-NP spatial configuration in optimizing Purcell enhancements. The results align with experimental observations, where CNP-doped scintillators consistently achieve higher enhancement factors than SNP-doped counterparts.

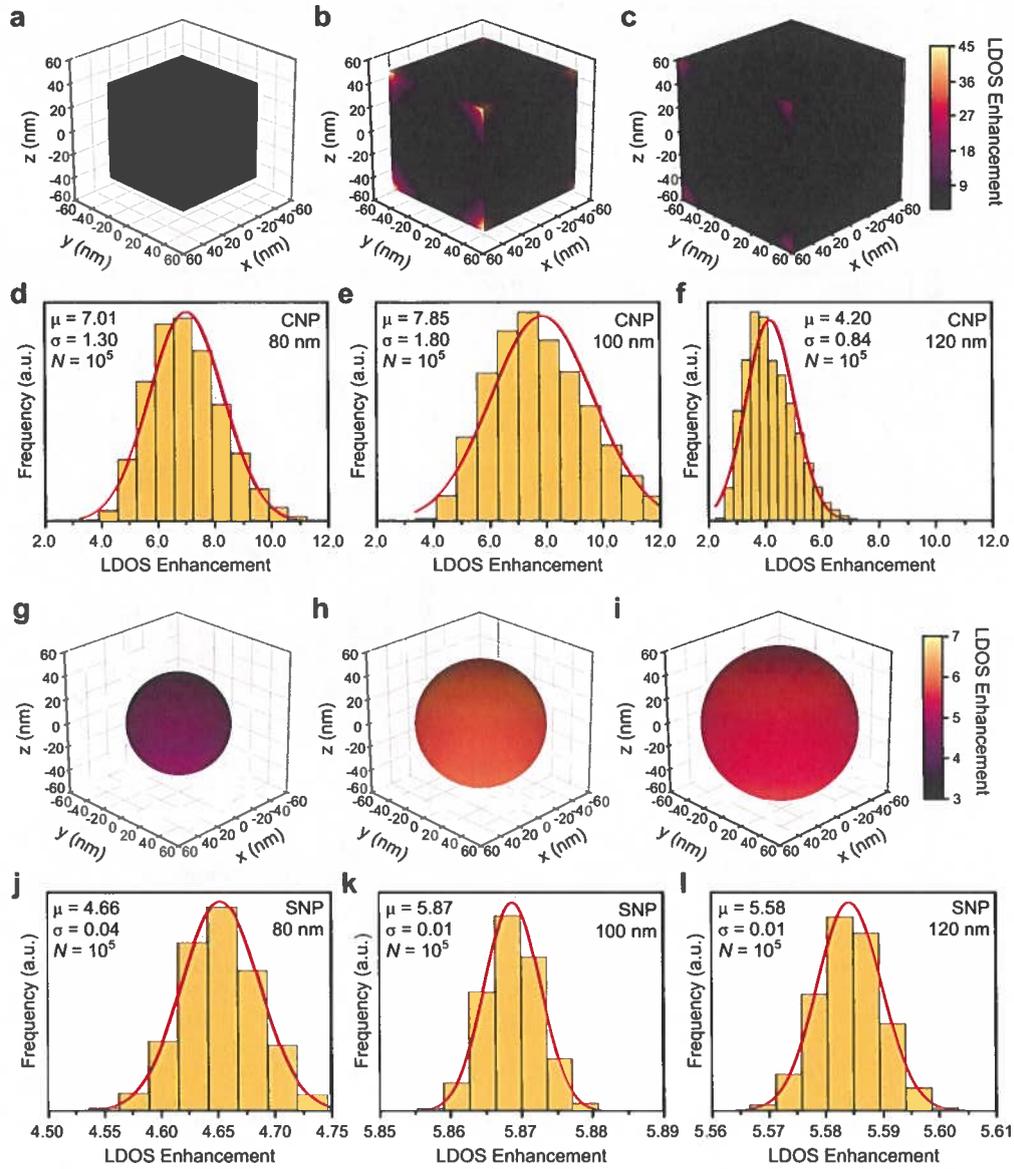

FIG. S9. **Influence of Ag NP size and shape on LDOS enhancement. a-c** and **g-i**, Three-dimensional maps of LDOS enhancements for CsPbBr$_3$ NCs doped with CNPs (**a-c**) and SNPs (**g-i**), respectively with corresponding histograms of LDOS enhancements (**d-f** and **j-l**).

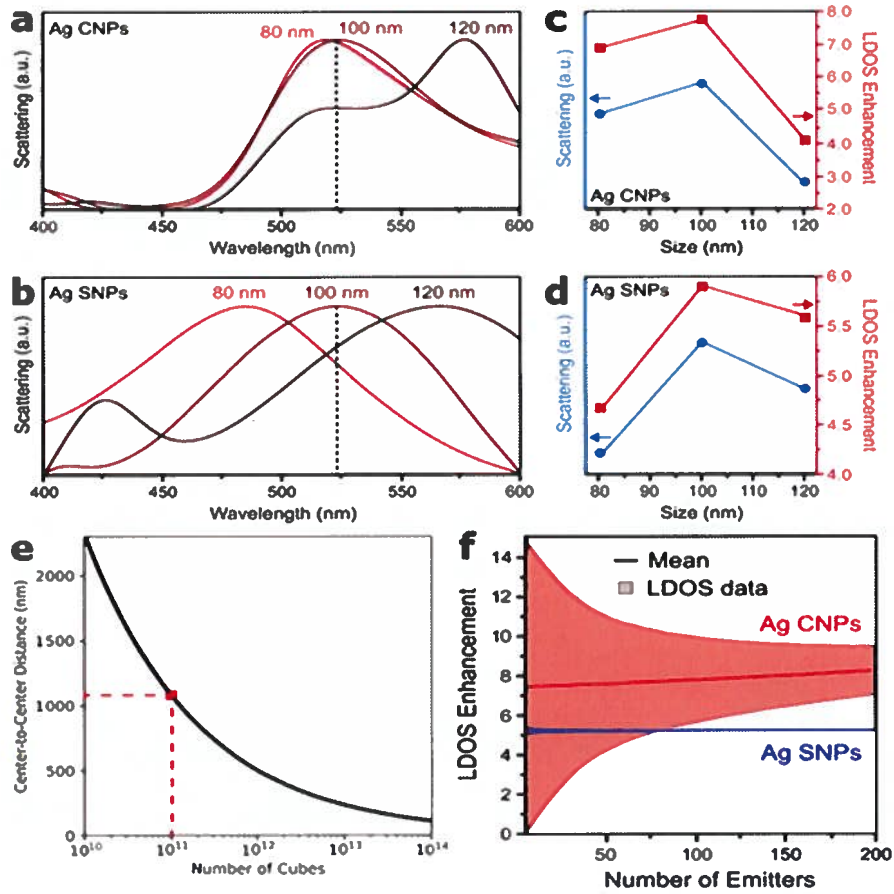

FIG. S10. **Influence of Ag NP size on its scattering spectra and FDTD simulations.** **a-d**, Influence of Ag NP size on the scattering spectra (**a** and **b**) and on LDOS enhancement (**c** and **d**), for Ag CNPs (**a** and **c**) and Ag SNPs (**b** and **d**). Blue points in **c** and **d** correspond to the fraction of maximum scattering spectra intensity at $CsPbBr_3$ emission wavelength. **e**, Center-to-center (C-C) distances between the NPs (100 nm-sized) in a 5×5×5 mm matrix. Red dashed lines correspond to the actual values achieved in examined samples. **f**, LDOS enhancement as a function of emitters attached to single Ag NP.

# TRANSMITTANCE

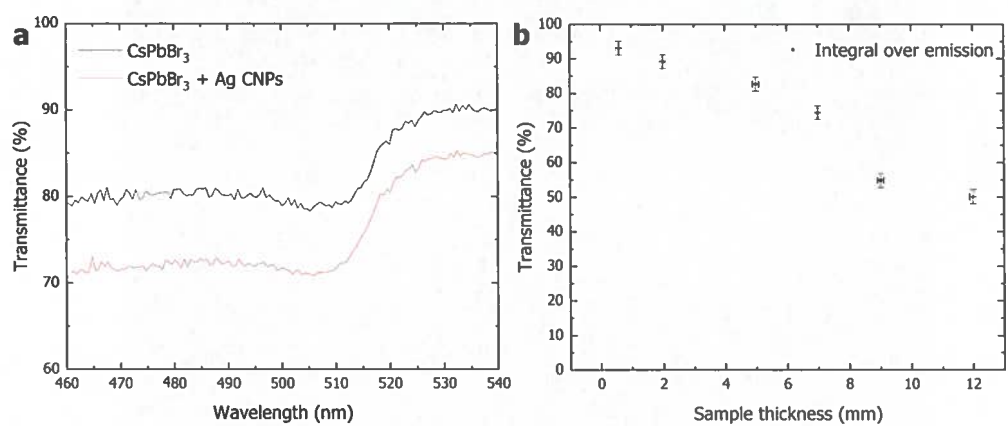

FIG. S11. **Transmittance.** a, Transmittance spectra of pure $CsPbBr_3$ and Ag CNP-doped samples, black and red curve, respectively. b, Transmittance as a function of pure $CsPbBr_3$ sample thickness. For proper evaluation, we present an integration result of transmittance over sample emission.

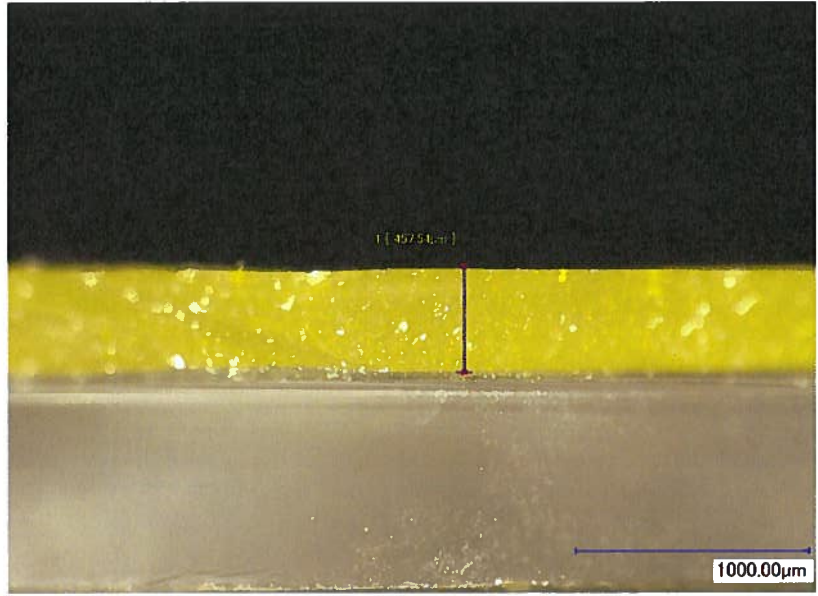

FIG. S12. **Samples for X-ray imaging.** Cross-section of CsPbBr$_3$ + Ag CNPs doped samples in a layer form, taken under the Keyence microscope.